\documentclass[twocolumn,twocolappendix]{aastex7}

\newcommand{\pj}[1]{#1}

\usepackage{graphicx}
\usepackage{amsmath}
\usepackage{siunitx}
\usepackage{enumitem}
\usepackage{multirow} 
\usepackage{booktabs}
\usepackage{placeins}
\usepackage{float}
\usepackage{caption}
\usepackage{subcaption}
\usepackage{longtable}
\usepackage{needspace}

\begin{document}

\title{Nine years of UVIT: assessing sensitivity variation}

\author[0009-0001-5407-3016]{Akanksha Dagore}
\affiliation{Indian Institute of Astrophysics, Koramangala II Block, Bangalore 560034, India}
\email[show]{akankshadagore@gmail.com}  

\author[0000-0003-1409-1903]{Prajwel Joseph}
\affiliation{Indian Institute of Astrophysics, Koramangala II Block, Bangalore 560034, India}
\email{prajwel.pj@gmail.com}

\author{S.N. Tandon}
\affiliation{Inter-University Center for Astronomy and Astrophysics, Pune 411007, India}
\email{sntandon@iucaa.in}  

\author[0000-0003-4612-620X]{Annapurni Subramaniam}
\affiliation{Indian Institute of Astrophysics, Koramangala II Block, Bangalore 560034, India}
\email{purni@iiap.res.in}  

\author{S.K. Ghosh}
\affiliation{Tata Institute of Fundamental Research, Mumbai 400005, India}
\email{swarna@tifr.res.in}  

\author[0000-0002-4998-1861]{C.S. Stalin}
\affiliation{Indian Institute of Astrophysics, Koramangala II Block, Bangalore 560034, India}
\email{stalin@iiap.res.in}  

\begin{abstract}

The Ultra-Violet Imaging Telescope (UVIT) is one of the five payloads onboard the first Indian multiwavelength astronomical observatory, \textit{AstroSat}, launched by the Indian Space Research Organisation on 28 September 2015. UVIT, designed for simultaneous imaging in the far-ultraviolet (FUV; 1300$-$1800 \AA) and near-ultraviolet (NUV; 2000$-$3000 \AA) channels, has completed nine years in orbit in 2024 despite the failure of the NUV channel in 2018. As the FUV optics is subject to possible reduction in sensitivity due to microscopic amounts of contaminants, we used the FUV data acquired by UVIT over the past nine years on the open cluster NGC 188 and the white dwarf HZ 4 to study sensitivity variations in the UVIT FUV channel. Our findings indicate no significant reduction in the sensitivity of the FUV channel over the last nine years, with no significant episodic variations due to unknown causes.

\end{abstract}



\section{Introduction} 
\label{sec:intro}

The Ultra-Violet Imaging Telescope (UVIT) is one of the five payloads onboard \textit{AstroSat} \citep{singh2014astrosat, agrawal2017astrosat}, India’s first space observatory targeted towards carrying out simultaneous multi-wavelength observations launched on 28 September 2015. \textit{AstroSat} is capable of observing in the ultraviolet (UV), visible (VIS), and X-ray spectral bands with the help of its five scientific payloads, namely UVIT, the Soft X-ray imaging Telescope (SXT), the Large Area X-ray Proportional Counter (LAXPC), the Cadmium Zinc Telluride Imager (CZTI), and the Scanning Sky Monitor (SSM). While the other four payloads observe in the X-ray window, UVIT performs simultaneous observations for a given region of the sky in far-ultraviolet (FUV; 1300$-$1800 \AA), near-ultraviolet (NUV; 2000$-$3000 \AA) and VIS (3200$-$5500 \AA) channels.

UVIT consists of twin Ritchey-Chr$\acute{\mathrm{e}}$tian telescopes, each with a primary and a secondary mirror of $\sim$375 mm and $\sim$140 mm diameter, respectively, along with a channel-filter configuration and a detector system. Each channel is equipped with a set of filters to select a narrower band-pass in which the intensified imaging detectors can operate in two distinct modes, namely high-gain photon counting (PC) mode and low-gain integration (INT) mode. The FUV and NUV channels perform observations in the photon counting mode, with a frame rate of $\sim29$ frames per second. The VIS channel, on the other hand, works in the low-gain integration mode, which gives out a reduced frame rate ($\sim1$ frame per second). The VIS channel is used to estimate the spacecraft pointing drift \citep{ghosh2021performance}. For the two UV channels, UVIT has a spatial resolution of $\sim\ang{;;1.5}$ FWHM and a field of view of $\sim\ang{;28}$ diameter. The C-MOS imagers of the detectors have a size of $512\times512$ pixels. Observations can also be carried out in already-defined, narrower window settings at faster frame rates. 

The expected mission duration for \textit{AstroSat} was planned to be five years \citep{singh2014astrosat}; however, as of September 2024, \textit{AstroSat} has completed nine years in orbit. The UVIT NUV channel became non-operational during this period on 30 March 2018. Despite the failure of the NUV channel, the FUV and VIS channels continue to function as intended. However, given its time in orbit, there exists a possibility that UVIT could have experienced a reduction in sensitivity. Possible factors can include degradation of reflectivity due to atomic oxygen \citep{banks2004low}, chemical contamination of the optical surfaces due to charged particles and UV radiation assisted reactions \citep{nahor1993degradation}, and accumulation of space dust and debris \citep{garoli2020mirrors}. Therefore, all these issues raise a need to investigate for any sensitivity degradation in UVIT.

In a previous study, \cite{tandon2020additional} found no significant degradation in the sensitivity of the FUV and NUV channels of UVIT, using stars in open cluster NGC 188 observations between 31 December 2015 and 23 October 2018. Since then, several years have passed, highlighting the need for new analysis. With the NUV channel non-functional since 2018, only the FUV channel requires sensitivity checks. To assess potential sensitivity variations in the FUV channel, frequent observations of a stable source spanning the lifetime of UVIT are necessary. The requirement of stability is obviously fulfilled by the white dwarf HZ 4, which is the calibration source for photometry with UVIT. However, unlike NGC 188, HZ 4 is not available for observation for all months of the year. Therefore, we use frequent observations of NGC 188 to monitor any significant variations, such as $>$5 per cent, on timescales of a few months, and infrequent observations of HZ 4 to monitor long-term decline at levels of one per cent. 
By comparing flux measurements of this source at various epochs, we can identify any changes in the FUV channel's sensitivity over time.

\begin{figure*}
    \centering
    \includegraphics[width=\textwidth]{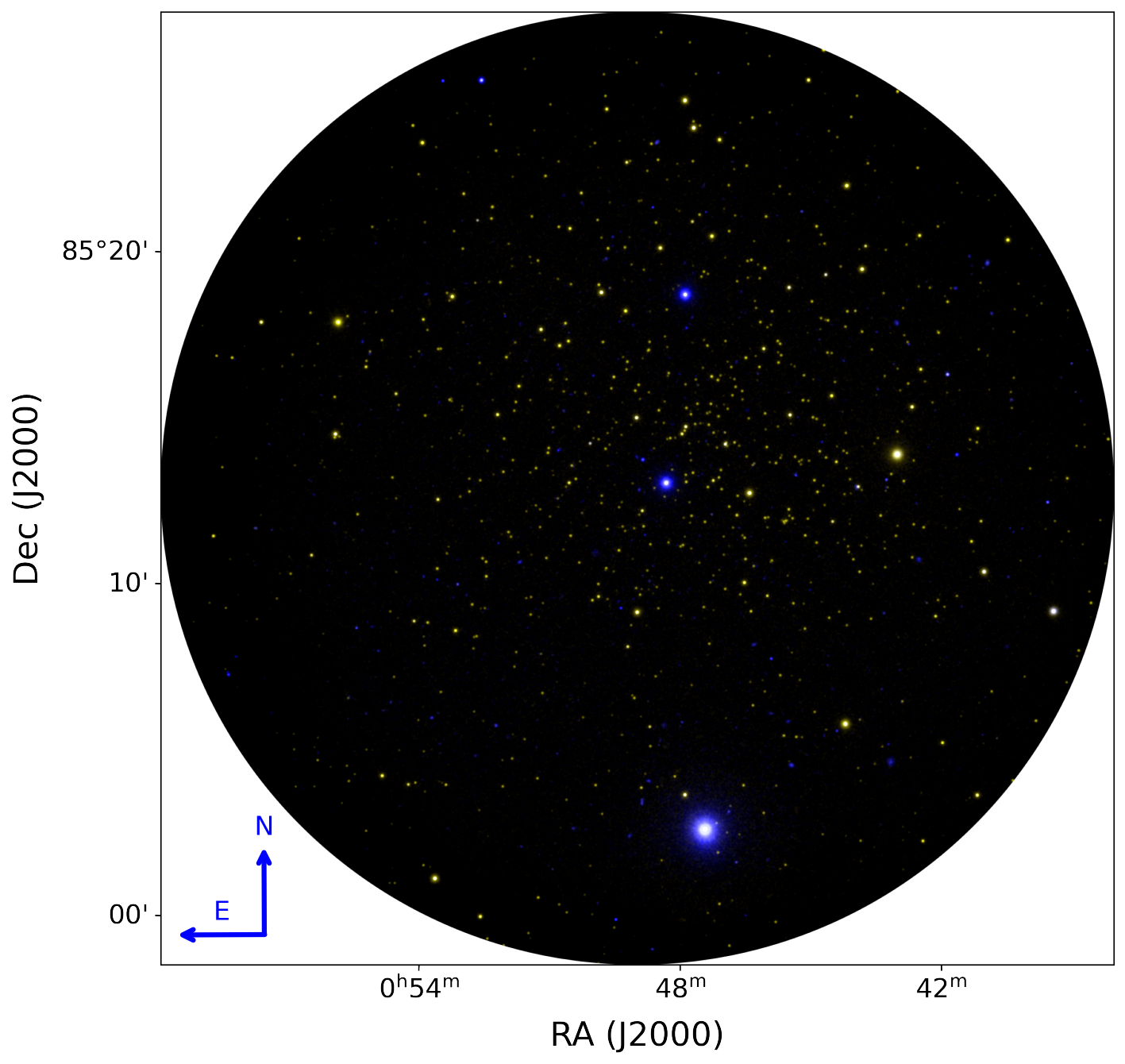}
    \caption{A pseudo-colour image of the NGC 188 field. The blue and yellow colours represent the observations conducted in the UVIT's FUV F148W \pj{(exposure time = 36484 seconds)} and NUV N279N \pj{(exposure time = 14954 seconds)} filters, respectively. The image has been processed to bring out the faint features.}
    \label{NGC_188 coloured}
\end{figure*}

UVIT captured its first sky image on 30 November 2015 of the open cluster NGC 188 ($\alpha = 00h \ 47m \ 11.5s$, $\delta$ = $\ang{+85}$$\ang{;14}$$\ang{;;38}$)\footnote{\url{https://simbad.u-strasbg.fr/simbad/sim-basic?Ident=NGC+188}} \citep{ghosh2021performance}, which has been monitored ever since for the telescope’s calibration and sensitivity studies, including the current one. NGC 188 observations are also conducted to test UVIT’s functionality after performing mitigation operations to resolve instrument problems.
\pj{Furthermore, these observations have contributed to scientific studies \citep{subramaniam2016hot, rani2021uocs}.}
Fig. \ref{NGC_188 coloured} illustrates a pseudo-colour image of the NGC 188 field created from the FUV F148W and NUV N279N filters of UVIT. While three FUV-bright sources are visible in Fig. \ref{NGC_188 coloured}, the brightest source suffers from saturation effects, and the other two were found to be variable during the course of this work. 
Nevertheless, this calibration source provides data to monitor any decline in sensitivity at levels below 1\% over long periods of years.


Section 2 discusses the characteristics of the observational data used for the analysis. The data reduction methodology is explained in Section 3. The analysis and results are given in Section 4. Lastly, the conclusion is reported in Section 5.


\section{Observational Data} 
\label{sec: Observational_data}

The UVIT observations are planned to coincide with the orbital phase of the satellite that falls within the Earth's shadow, which blocks direct exposure from the Sun to prevent instrument damage. \textit{AstroSat} covers a complete \textit{orbit} around the Earth in $\sim$100 minutes, and any uninterrupted imaging session during this phase with a specific filter-window configuration is termed an \textit{episode}.

The observational data for HZ 4 and NGC 188 were obtained from the \textit{AstroSat} Archive of the Indian Space Science Data Centre (ISSDC)\footnote{\url{https://astrobrowse.issdc.gov.in/astro\_archive/archive/Home.jsp}}. The data were downloaded in the science-ready Level2 (L2) format, which was processed by the L2 pipeline software (version 6.3; \citealt{ghosh2021performance, ghosh2022automated}) at the UVIT Payload Operation Centre (POC) and transferred to ISSDC for archiving and dissemination. The L2 pipeline performs a series of operations on the Level1 (L1) data and refines it into science-ready L2 data products. 

Depending on the duration of the observation, the imaging sessions may be divided into episodes that span multiple orbits. The 6.3 version L2 data generated for each episode is organised and stored separately for each channel and usually consists of the following elements in FITS format:
\begin{itemize}[wide, labelwidth=!,itemindent=!,labelindent=0pt, leftmargin=0em, noitemsep]
    \item{image file in the instrument coordinate system ($x$, $y$)};
    \item{image file in the astronomical coordinate system ($\alpha$, $\delta$)};
    \item{exposure map in the astronomical coordinate system ($\alpha$, $\delta$)};
    \item{error/noise map in the astronomical coordinate system ($\alpha$, $\delta$)}; and
    \item{a tabulated list containing information about the recorded photon events.}
\end{itemize}
For the current study, 4 UVIT observations of HZ 4 and 35 of the NGC 188 field were analysed. They span from 22 March 2016 to 28 August 2024. Many of the individual observations consist of multiple episodes, leading to 13 episode images for HZ 4 and 57 for NGC 188. We have used only FUV F148W filter imaging data and discarded those episodes with less than 130 seconds of exposure time. While the NGC 188 observations were obtained in the $512\times512$ pixels window mode, the HZ 4 observations were in the $200\times200$ pixels window mode. An observation taken in $200\times200$ pixels window mode provides a higher frame rate ($\approx$ 180 frames per second) as compared to that with the window of $512\times512$ pixels ($\approx$ 29 frames per second), and hence reduces saturation effects. Tables \ref{table:obs_info_hz4} and \ref{table:obs_info_ngc_188} list the observation IDs, observation dates, number of episode images, and exposure times for the HZ 4 and NGC 188 observations, respectively. Regardless of the window mode, all image products have a size of $4800\times4800$ sub-pixels (1 sub-pixel length $=$ $1/8$ detector pixel length).

\begin{deluxetable*}{ccccc}
\tablecaption{The table lists the HZ 4 observation IDs that were obtained from ISSDC along with their respective dates of observation, number of episodes, and exposure time in seconds.
\label{table:obs_info_hz4}}
\tablehead{
\colhead{No.} & \colhead{Observation ID} & \colhead{Date of observation} &  \colhead{No. of episode images} & \colhead{Exposure time} \\
\colhead{} & \colhead{} & \colhead{(dd-mm-yyyy)} & \colhead{} & \colhead{(s)} }
\startdata
\multirow{3}{*}{1.} & \multirow{3}{*}{C02\textunderscore002T01\textunderscore9000000888} & \multirow{3}{*}{16-12-2016} & \multirow{3}{*}{3} & 375 \\
& & & & 375 \\ 
& & & & 375 \\ \hline
\multirow{4}{*}{2.} & \multirow{4}{*}{C03\textunderscore013T01\textunderscore9000001586} & \multirow{4}{*}{06-10-2017} & \multirow{4}{*}{4} & 336 \\
& & & & 374 \\ 
& & & & 261 \\
& & & & 245 \\ \hline
\multirow{4}{*}{3.} & \multirow{4}{*}{C04\textunderscore010T01\textunderscore9000003158} & \multirow{4}{*}{12-09-2019} & \multirow{4}{*}{4} & 358 \\
& & & & 358 \\ 
& & & & 271 \\
& & & & 164 \\ \hline
\multirow{2}{*}{4.} & \multirow{2}{*}{C02\textunderscore002T01\textunderscore9000006092} & \multirow{2}{*}{24-02-2024} & \multirow{2}{*}{2} & 359 \\
& & & & 359 \\
\enddata
\end{deluxetable*}

\startlongtable
\begin{deluxetable*}{ccccc}
\tablecaption{The table lists the NGC 188 observation IDs that were obtained from ISSDC along with their respective dates of observation, number of episodes, and exposure time in seconds.
\label{table:obs_info_ngc_188}}
\tablehead{
\colhead{No.} & \colhead{Observation ID} & \colhead{Date of observation} &  \colhead{No. of episode images} & \colhead{Exposure time} \\
\colhead{} & \colhead{} & \colhead{(dd-mm-yyyy)} & \colhead{} & \colhead{(s)}
}
\startdata
\multirow{2}{*}{1.} & \multirow{2}{*}{T01\textunderscore034T01\textunderscore9000000392} & \multirow{2}{*}{22-03-2016} & \multirow{2}{*}{2} & 562 \\
& & & & 466 \\ \hline
2. & G05\textunderscore209T01\textunderscore9000000458 & 18-05-2016 & 1 & 1105 \\ \hline
3. & G05\textunderscore258T01\textunderscore9000000666 & 15-09-2016 & 1 & 621 \\ \hline
4. & C02\textunderscore016T01\textunderscore9000000992 & 30-01-2017 & 1 & 1195 \\ \hline
\multirow{3}{*}{5.} & \multirow{3}{*}{C02\textunderscore030T01\textunderscore9000001168} & \multirow{3}{*}{16-04-2017} & \multirow{3}{*}{3} & 307 \\
& & & & 433 \\
& & & & 352 \\ \hline
6. & C03\textunderscore015T01\textunderscore9000001788 & 21-12-2017 & 1 & 149 \\ \hline
\multirow{2}{*}{7.} & \multirow{2}{*}{T02\textunderscore002T01\textunderscore9000001914} & \multirow{2}{*}{23-02-2018} & \multirow{2}{*}{2} & 1855 \\
& & & & 1039 \\ \hline
8. & C03\textunderscore015T02\textunderscore9000002008 & 04-04-2018 & 1 & 431 \\ \hline
9. & C03\textunderscore015T04\textunderscore9000002240 & 20-07-2018 & 1 & 1063 \\ \hline
\multirow{3}{*}{10.} & \multirow{3}{*}{C03\textunderscore015T04\textunderscore9000002328} & \multirow{3}{*}{26-08-2018} & \multirow{3}{*}{3} & 255 \\
& & & & 336 \\ 
& & & & 541 \\ \hline
\pagebreak
\multirow{3}{*}{11.} & \multirow{3}{*}{C03\textunderscore015T03\textunderscore9000002366} & \multirow{3}{*}{14-09-2018} & \multirow{3}{*}{3} & 554 \\
& & & & 424 \\ 
& & & & 155 \\ \hline
12. & C03\textunderscore015T04\textunderscore9000002382 & 21-09-2018 & 1 & 1140 \\ \hline
13. & C04\textunderscore009T01\textunderscore9000002448 & 23-10-2018 & 1 & 1140 \\ \hline
14. & C04\textunderscore009T02\textunderscore9000002618 & 04-01-2019 & 1 & 1140 \\ \hline
15. & C05\textunderscore013T02\textunderscore9000003364 & 13-12-2019 & 1 & 1140 \\ \hline
16. & C05\textunderscore013T03\textunderscore9000003376 & 19-12-2019 & 1 & 508 \\ \hline
17. & C05\textunderscore013T04\textunderscore9000003660 & 12-05-2020 & 1 & 1144 \\ \hline
\multirow{2}{*}{18.} & \multirow{2}{*}{T03\textunderscore247T01\textunderscore9000003914} & \multirow{2}{*}{05-10-2020} & \multirow{2}{*}{2} & 193 \\
& & & & 546 \\ \hline
19. & C05\textunderscore013T06\textunderscore9000003920 & 15-10-2020 & 1 & 736 \\ \hline
20. & C06\textunderscore010T01\textunderscore9000004504 & 03-07-2021 & 1 & 678 \\ \hline
\multirow{2}{*}{21.} & \multirow{2}{*}{C06\textunderscore010T02\textunderscore9000004618} & \multirow{2}{*}{02-08-2021} & \multirow{2}{*}{2} & 1469 \\
& & & & 875 \\ \hline
\multirow{2}{*}{22.} & \multirow{2}{*}{C06\textunderscore010T03\textunderscore9000004672} & \multirow{2}{*}{27-08-2021} & \multirow{2}{*}{2} & 991 \\ 
& & & & 1480 \\ \hline
\multirow{2}{*}{23.} & \multirow{2}{*}{C06\textunderscore010T05\textunderscore9000004736} & \multirow{2}{*}{25-10-2021} & \multirow{2}{*}{2} & 1558 \\ 
& & & & 204 \\ \hline
24. & T04\textunderscore065T01\textunderscore9000004750 & 01-11-2021 & 1 & 163 \\ \hline
\multirow{3}{*}{25.} & \multirow{3}{*}{C06\textunderscore010T05\textunderscore9000005424} & \multirow{3}{*}{15-12-2022} & \multirow{3}{*}{3} & 482 \\
& & & & 1319 \\ 
& & & & 985 \\ \hline
\multirow{2}{*}{26.} & \multirow{2}{*}{C06\textunderscore010T05\textunderscore9000005436} & \multirow{2}{*}{23-12-2022} & \multirow{2}{*}{2} & 1046 \\ 
& & & & 1732 \\ \hline
\multirow{2}{*}{27.} & \multirow{2}{*}{C06\textunderscore010T05\textunderscore9000005484} & \multirow{2}{*}{15-01-2023} & \multirow{2}{*}{2} & 1435 \\ 
& & & & 1174 \\ \hline
\multirow{2}{*}{28.} & \multirow{2}{*}{T05\textunderscore101T01\textunderscore9000005578} & 22-04-2023 & \multirow{2}{*}{2} & 793 \\ 
& & 23-04-2023 & & 212 \\ \hline
\multirow{2}{*}{29.} & \multirow{2}{*}{T05\textunderscore101T01\textunderscore9000005586} & 28-04-2023 & \multirow{2}{*}{2} & 851 \\ 
& & 29-04-2023 & & 182 \\ \hline
\multirow{2}{*}{30.} & \multirow{2}{*}{C06\textunderscore010T05\textunderscore9000005670} & \multirow{2}{*}{30-05-2023} & \multirow{2}{*}{2} & 522 \\ 
& & & & 890 \\ \hline
\multirow{2}{*}{31.} & \multirow{2}{*}{C06\textunderscore010T01\textunderscore9000005800} & \multirow{2}{*}{29-07-2023} & \multirow{2}{*}{2} & 1269 \\ 
& & & & 227 \\ \hline
\multirow{2}{*}{32.} & \multirow{2}{*}{C06\textunderscore010T01\textunderscore9000005816} & \multirow{2}{*}{13-08-2023} & \multirow{2}{*}{2} & 195 \\ 
& & & & 390 \\ \hline
33. & C09\textunderscore011T02\textunderscore9000006310 & 24-06-2024 & 1 & 736 \\ \hline
\pagebreak
\multirow{2}{*}{34.} & \multirow{2}{*}{C09\textunderscore011T03\textunderscore9000006410} & \multirow{2}{*}{25-08-2024} & \multirow{2}{*}{2} & 732 \\ 
& & & & 1379 \\ \hline
35. & C09\textunderscore011T01\textunderscore9000006418 & 28-08-2024 & 1 & 573 \\
\enddata
\end{deluxetable*}


\section{Methodology} 
\label{methodology}

We need to obtain the fluxes of select sources at different epochs to assess the FUV channel sensitivity variations. This section explains the steps followed to obtain source fluxes from the episode images.

\subsection{Source selection}

\begin{figure}
    \centering
    \includegraphics[width=\columnwidth]{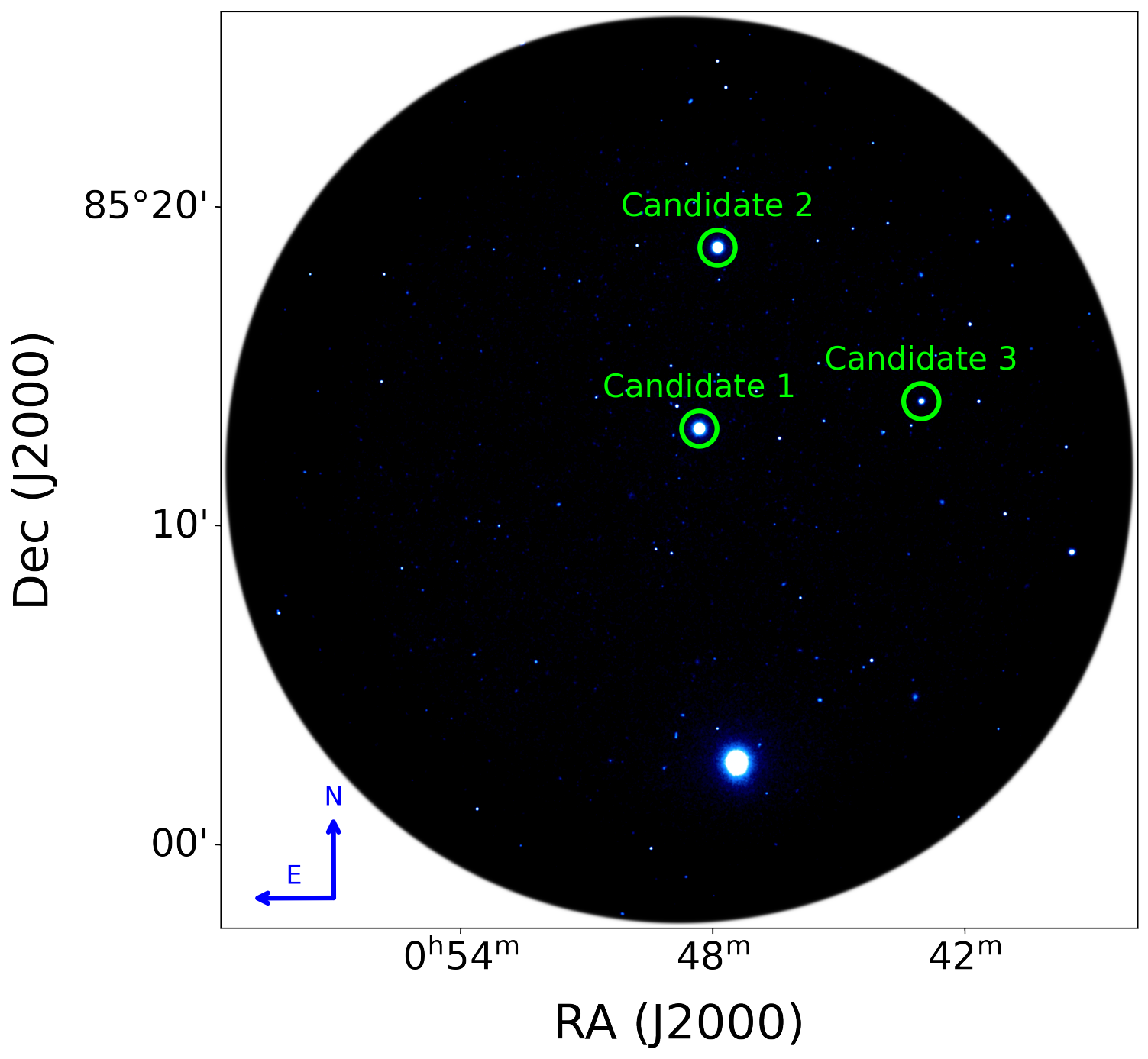}
    \caption{The three sources (green circles) selected in the NGC 188 field to carry out UVIT sensitivity checks are marked in the UVIT F148W image. The image has been processed to bring out the faint features.}
    \label{target_candidates}
\end{figure}

The HZ 4 source was chosen to carry out sensitivity checks in the HZ 4 field. In all the exposures, HZ 4 remains close to the centre of the field, and hence the data are minimally affected by errors in the flat-field correction. We chose three sources in the NGC 188 field. 
We ensured that the selected sources in the NGC 188 field did not lie outside of an 1800 sub-pixel radius from the image centroid, were not saturated, and, most importantly, were present in all 57 episode images. Fig. \ref{target_candidates} shows the selected sources in the NGC 188 field. The brightest source in the field was saturated, so we chose the next three brightest sources that met our criteria. We identified these three sources across all 57 episode images using the \texttt{Aafitrans} Python package \citep{prajwel_aafitrans, BEROIZ2020100384} which matches centroids of the detected sources in a reference image against all remaining images (see Appendix \ref{ssec: source_detection} for details on the source detection method). The sources are located at different radii in the field and, as the field rotates over the year, they explore different parts of it. For this reason, the corresponding data are affected by any errors in the flat-field correction.

\subsection{Aperture photometry}
\label{ssec: aperture_photometry}

A key step in performing aperture photometry is the subtraction of the background. The method used to generate the background maps for episode images is explained in Appendix \ref{appendix: background_estimation}. Each sub-pixel in the background subtracted image of an individual episode gives a measure of the photons recorded by the detector in counts per second (CPS) units. The aperture photometry was done using the \texttt{Photutils} package \citep{larry_bradley_2022_6825092}. We used an aperture of 12 sub-pixels radius that encompasses $88.6\%$ of the source encircled energy in the FUV channel \citep{tandon2020additional}. 

\subsection{Applying corrections to the selected source fluxes}
\label{ssec:additioanl_corrections}

There are factors that can influence the calculated flux values and, therefore, must be accounted for. These arise due to the selected aperture size and the detector's inability to distinguish multiple photons falling within a small region of $3\times3$ detector pixels in a single frame (saturation effects). We applied aperture and saturation corrections to address them, respectively.

A method to correct for the saturation effects was put forward by \cite{tandon2017orbit} based on the idea that the incident photons falling on the detector follow a Poisson distribution. However, this method is specifically designed for source flux in CPS without flat-field correction. As the episode images have flat-field corrections, it is necessary first to reverse the flat-field correction and extract the non-flat-fielded flux in CPS before applying aperture and saturation corrections.  Therefore, we applied the corrections as follows:
\begin{enumerate}[label=(\roman*), wide, labelwidth=!,itemindent=!,labelindent=0pt, leftmargin=0em, noitemsep]
    \item removed flat-field corrections from the flux in CPS to get the non-flat-fielded flux in CPS;
    \item applied aperture correction;
    \item applied saturation correction; and
    \item reintroduced the flat-field correction to obtain the final flat-fielded flux in CPS with aperture and saturation corrections.
\end{enumerate}
The methodology for these corrections is explained in the following sections.

\subsubsection{Flat-field correction removal}
\label{subsec: FF}

The flat-fields for each filter are stored in the calibration data files, which are used by the L2 pipeline to generate L2 data. Since the L2 pipeline applies the flat-field correction before the telescope pointing drift correction, removing the flat-field correction from L2 images is not straightforward. Therefore, we adopted the following approximate approach to remove the flat-field correction.

The UVIT images have a dimension of $4800\times4800$ sub-pixels, while the flat-fields are $512\times512$ pixels \citep{ghosh2022automated}. For each source coordinate ($x_{s}$, $y_{s}$) in the instrument coordinate system image, the corresponding location in the flat-field ($x_{f}, y_{f}$) was calculated using the equations:
\begin{align}
    x_{f} = 256 + \frac{x_{s} - x_{centre}}{8} , \\ 
    y_{f} = 256 + \frac{y_{s} - y_{centre}}{8}.
\end{align}

Here, ($x_{centre}$, $y_{centre}$) denotes the coordinate of the field centre, which serves as the reference point for mapping the image to the flat-field grid.
A $21\times21$ pixel window was placed centred on ($x_{f}, y_{f}$) in the flat-field, and the average of the pixel values within this window ($FF_{avg}$) was estimated. Then, the non-flat-fielded flux in CPS ($\mathrm{CPS}_{noFF}$) was obtained using the equation,
\begin{align}
    \mathrm{CPS}_{noFF} = \frac{\mathrm{CPS}_{measured} - \mathrm{CPS}_{bkg}}{FF_{avg}} \ .
\end{align}
where $\mathrm{CPS}_{measured}$ is the measured source flux in CPS.
Background maps were generated for each UVIT image (see Appendix~\ref{appendix: background_estimation}) and used to obtain the local background level, $\mathrm{CPS}_{bkg}$.

The uncertainty in the measured flux, $\mathrm{CPS}_{measured\_error}$, was calculated using the following equation:
\begin{equation}
    \mathrm{CPS}_{measured\_error} = \frac{\sqrt{(\mathrm{CPS}_{measured} / FF_{avg} ) \times t_{exp}}}{t_{exp}} \ ,
\end{equation}
where $t_{exp}$ is the exposure time in seconds for a given observation. 
Uncertainties propagated through each correction step were evaluated using $\mathrm{CPS}_{measured\_error}$.

\subsubsection{Aperture correction} 

\cite{tandon2020additional} provides a table listing the point source encircled energy percentages as a function of the aperture radius in sub-pixels for both the NUV and FUV channels. From this table, we derived the aperture correction factor ($f_{ac}$) to be 100/88.6 for a 12 sub-pixel radius aperture in the FUV channel. The aperture corrected flux in CPS ($\mathrm{CPS}_{ac}$) was then calculated using the equation:
\begin{align}
    \mathrm{CPS}_{ac} = \mathrm{CPS}_{noFF} \times f_{ac} \ .
\end{align}

\subsubsection{Saturation correction}

When multiple photons fall within a small area of $3\times3$ UVIT detector pixels during a single frame in PC mode, the detector registers them as a single photon. This leads to a saturation of the measured photon count rate. Unless the source counts per frame (CPF) $<<1$, the saturation effect must be corrected while estimating the photon count rate, as failing to do so would underestimate the source flux. The UVIT saturation correction method for observed source CPF up to 0.6 is given by \cite{tandon2017orbit} and is elaborated below.

\begin{figure}
    \centering
    \includegraphics[width=\columnwidth]{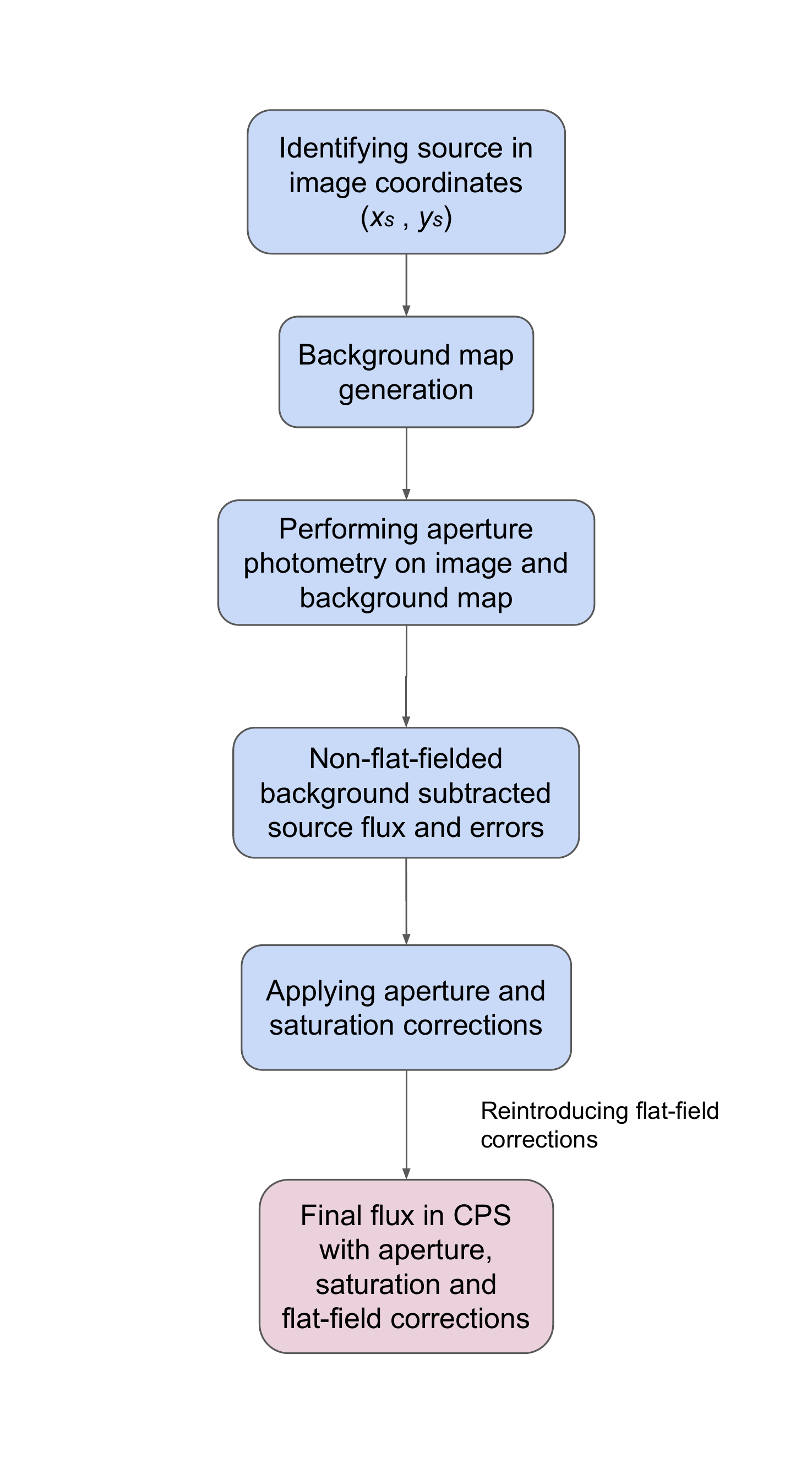}
    \caption{
    All steps followed to get the corrected flux for a point source.
    }
    \label{flowchart}
\end{figure}

In UVIT images, the central $40\times40$ sub-pixels region of a source contains $97\%$ of the source photons and is more susceptible to saturation effects than the low-count pedestal of its PSF profile. Therefore, we need to apply saturation correction within this region. If CPF5 denotes $97\%$ of the observed CPF, and ICPF5 represents the corresponding actual counts per frame using Poisson statistics, then CPF5 and ICPF5 can be related using the following equations:
\begin{align}
    \mathrm{CPF5} & = 1 - e^{-\mathrm{ICPF5}} \ ,  \\
    \mathrm{ICPF5} & = -\mathrm{ln} (1-\mathrm{CPF5}) \ .
\end{align}
The difference between CPF5 and ICPF5 should give the ideal correction for saturation, i.e., 
\begin{align}
    \mathrm{ICORR} = \mathrm{ICPF5} - \mathrm{CPF5}  \ . \label{eq:ICORR}
\end{align}
However, the real correction for saturation, RCORR, is found to be different from the ideal correction, ICORR. \cite{tandon2017orbit} provides a relation between RCORR and ICORR as follows:
\begin{align}
    \mathrm{RCORR} = \mathrm{ICORR} \times (0.89 - 0.30 \times (\mathrm{ICORR})^{2} )  \ . \label{eq.RCORR}
\end{align}
We used the above prescription to correct $\mathrm{CPS}_{ac}$ for saturation. We obtained CPF5 from $\mathrm{CPS}_{ac}$ using the equations,
\begin{align}
    \mathrm{CPF}_{ac} & = \frac{\mathrm{CPS}_{ac}}{\mathrm{frame\ rate}} \ , \\
    \mathrm{CPF5} & = \mathrm{CPF}_{ac} \times 0.97 \ .   
\end{align}
Then, RCORR was calculated using CPF5. The saturation corrected flux in CPS ($\mathrm{CPS}_{sc}$) was obtained as,
\begin{align}
\mathrm{CPS}_{sc} = \mathrm{CPS}_{ac} + (\mathrm{RCORR} \times \mathrm{frame\ rate}) \ .
\end{align}
We have now obtained the non-flat-fielded flux in CPS with aperture and saturation corrections ($\mathrm{CPS}_{sc}$). Finally, we reintroduce the flat-field correction:
\begin{align}
    \mathrm{CPS}_{final} & = \mathrm{CPS}_{sc} \times FF_{avg} \ ,
\end{align}
where $\mathrm{CPS}_{final}$ is the final corrected flux in CPS for a given source. A flow chart of the steps adopted to get the final fluxes of the point sources is given in Fig. \ref{flowchart}.

\section{Sensitivity analysis}
\label{sec:sens_analysis}

Here we analyse the multi-epoch point source fluxes from the HZ 4 and NGC 188 fields to investigate potential variations in UVIT sensitivity. The source fluxes were calculated following the methodology described in Section \ref{methodology}.

\subsection{HZ 4}
\begin{figure}
    \centering
    \includegraphics[width=\columnwidth]{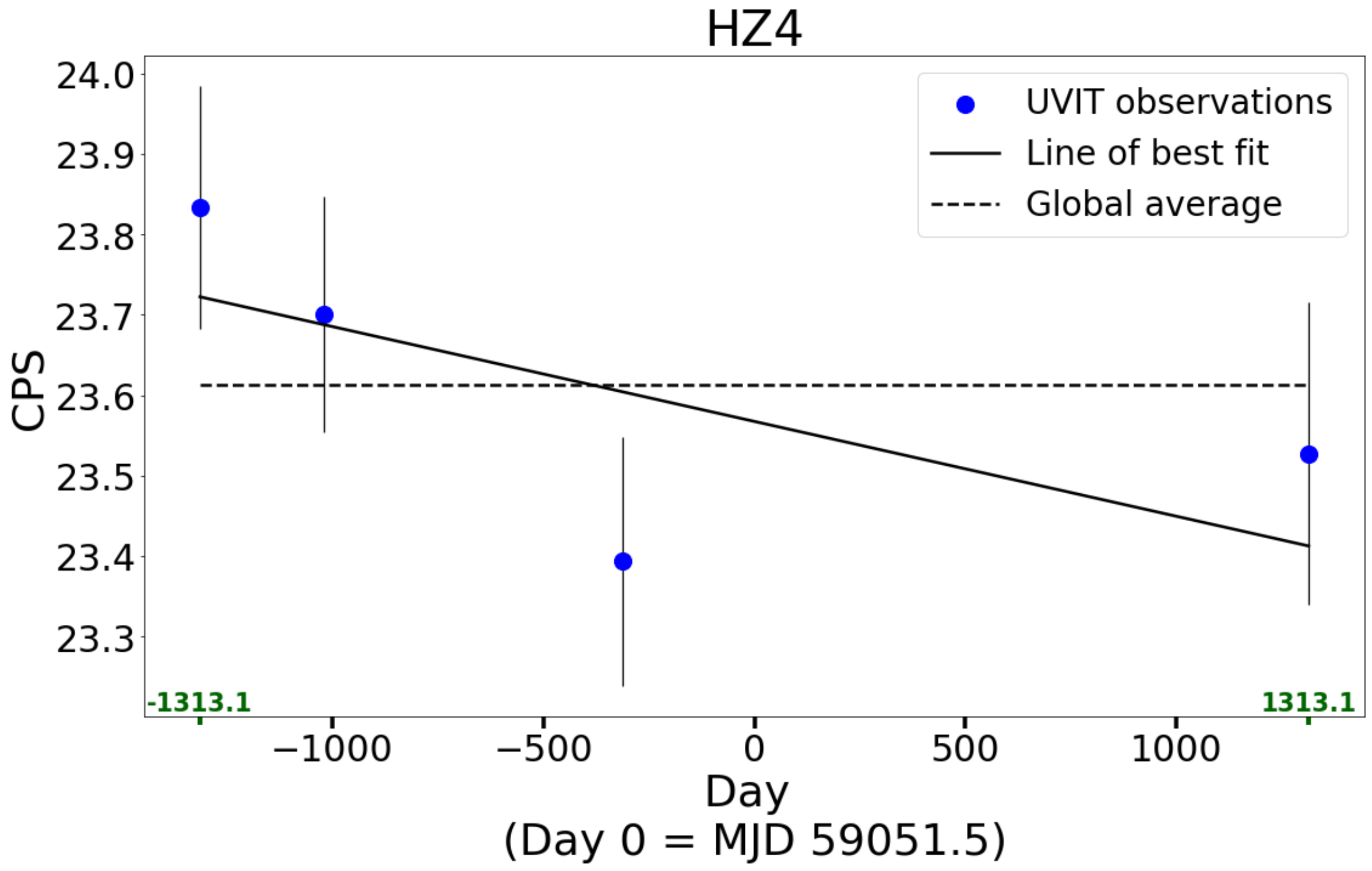}
    \caption{The FUV light curve for HZ4. The solid and dashed lines mark the best fit line and the global average of the data, respectively. Green minor ticks and labels mark the times of the first and last observations.}
    \label{HZ4_combined}
\end{figure}

To analyse the data, we calculated the mean source flux for each observation ID by averaging the episode-wise source fluxes within that ID (see Table \ref{table:flux_err_HZ4} listing the corrected flux measurements and associated uncertainties). The resulting light curve along with its best-fit line and global average are shown in Fig. \ref{HZ4_combined}, with the line fit statistics provided in Table \ref{table:line_fit_stats_HZ4}.
 The linear least squares fits were weighted by the inverse square of the measurement uncertainties (i.e. 1 / $\mathrm{CPS}_{final\_error}^2$).
The intercept was determined at the midpoint of the observation time period and is marked as Day 0. 
This approach is adopted to compute the intercepts in all subsequent analyses presented in this paper.
The best fit line equation for HZ 4 is given by (\ref{hz4_uvit}). In the equation, the independent variable, MJD, refers to the Modified Julian Date, while CPS(MJD) represents CPS as a function of MJD.
\begin{align}
    \text{CPS(MJD)} = & \ (\num{-1e-04} \pm \num{1e-04})(\text{MJD} - 59051.5) \nonumber \\
    & + \ 23.6 \pm 0.1 \ . \label{hz4_uvit}
\end{align}
Although the estimated slope is negative, the large associated error introduces significant uncertainty. It is also seen that a constant value gives an acceptable fit, and we infer that the sensitivity shows no degradation. We can also put a 95\% confidence upper limit on the degradation over the 9 years as 4\%.





\begin{deluxetable*}{cccccccc}
\tablecaption{Line fit statistics for the HZ4 FUV data shown in Fig. \ref{HZ4_combined}.
\label{table:line_fit_stats_HZ4}}
\tablehead{
\colhead{Source} & \colhead{MJD at Day 0} & \multicolumn{2}{c}{Slope} & \multicolumn{2}{c}{Intercept at Day 0} & \multicolumn{2}{c}{Global average} \\
\colhead{} & \colhead{} & \multicolumn{2}{c}{(CPS/MJD)} & \multicolumn{2}{c}{(CPS)} & \multicolumn{2}{c}{(CPS)} \\
\cline{3-8}
\colhead{} & \colhead{} & \colhead{Value} & \colhead{Error} & \colhead{Value} & \colhead{Error} & \colhead{Value} & \colhead{Error}
}
\startdata
HZ 4 & 59051.5 & -1e-04 & 1e-04 & 23.6 & 0.1 & 23.61 & 0.08 \\
\enddata
\end{deluxetable*}

\subsection{NGC 188}
\begin{figure*}
    \centering
    \includegraphics[width=\textwidth]{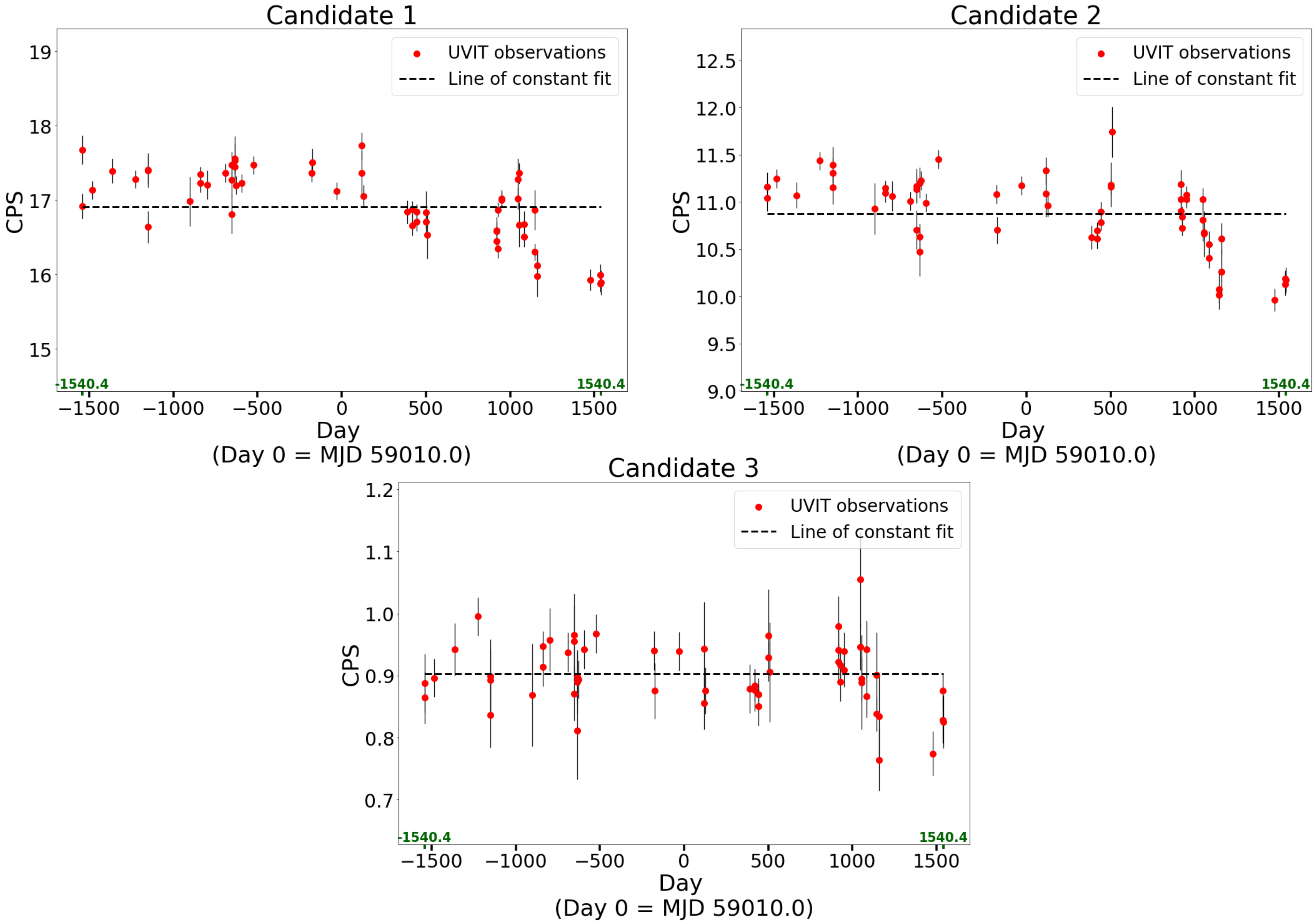}
    \caption{The FUV light curves for the selected three sources in the field of NGC 188. The dashed line marks the line of constant fit. Green minor ticks and labels mark the times of the first and last observations.}
    \label{line_fit}
\end{figure*}

The light curves of the three selected sources in the NGC 188 field are shown in Fig. \ref{line_fit} (see Table \ref{table:flux_err_NGC188} listing the corrected flux measurements and associated uncertainties). From this figure, it can be observed that candidates 1 and 2 are relatively bright compared to candidate 3. All three show flux variations, particularly candidates 1 and 2 which exhibit significant changes exceeding the associated uncertainties. 
As the data on HZ 4 is consistent with no sensitivity degradation, we assume the same holds true for these sources. Any signal variations are, therefore, attributed to errors in flat-field correction, temporal variations in the sources themselves, or episodic unexplained variations in the FUV sensitivity. 
If episodic FUV sensitivity variations of unknown origin exist, the observed variability can be interpreted as an upper limit on their amplitude.
It is also possible that the variability is intrinsic to the sources themselves; however, a detailed investigation of this is beyond the scope of the present work and will be addressed separately.

\begin{deluxetable*}{cccc}
\tablecaption{Weighted RMSD values from the line of constant fit for the FUV data of the three selected sources from NGC 188 shown in Fig. \ref{line_fit}.
\label{table:line_fit_stats}}
\tablehead{
\colhead{Source} & \multicolumn{3}{c}{Line of constant fit}\\
\colhead{} & \multicolumn{3}{c}{(CPS)} \\
\cline{2-4}
\colhead{} & \colhead{Value} & \colhead{Error} & \colhead{RMSD}
}
\startdata
Candidate 1 & 16.91 & 0.06 & 0.463 (2.74\%)\\ 
Candidate 2 & 10.88 & 0.05 & 0.376 (3.46\%)\\
Candidate 3 & 0.902 & 0.006 & 0.046 (5.12\%)\\
\enddata
\end{deluxetable*}

Using \texttt{Astrometry.net} software (\citealt{lang2010astrometry}; see also Appendix \ref{appendix: index_fils}) and the SIMBAD Astronomical Database\footnote{\url{https://simbad.u-strasbg.fr/simbad/}} \citep{wenger2000simbad}, candidate 1 was identified as \pj{a blue straggler star with a post-AGB/HB companion (WOCS-5885) by \cite{subramaniam2016hot}, while candidate 2 was found to be a subdwarf (WOCS-4918,  \citealt{rani2021uocs})}. 
Regarding the possibility of any episodic variations in the sensitivity due to any possible unknown cause, these results indicate that these are not more than $\sim$5\%. To address this, we computed the weighted root mean square deviations (RMSD) from the line of constant fit for the NGC 188 observations, which are provided in Table \ref{table:line_fit_stats}.

\section{Summary}

In September 2024, UVIT achieved a significant milestone by completing nine years in orbit. As the telescope ages, it is essential to assess whether any sensitivity degradation has occurred due to various potential factors. To investigate this, we analysed the multi-epoch data from white dwarf HZ 4, the calibration source for UVIT. We also analysed the frequent observations of the open cluster NGC 188 to check for any short-term episodic variations in sensitivity.

The results for HZ 4 are consistent with no degradation in sensitivity at the centre of the field for the FUV channel. They also provide a 95\% confidence upper limit of $\sim$4\% on degradation over the full period of 9 years. Based on these findings, we conclude that there is no significant degradation of sensitivity in the FUV channel. Since the sensitivity of the NUV channel is far less likely to degrade compared to the FUV channel, we infer that its sensitivity remained stable throughout its operation until 2018. As far as any possibility of a shorter-term sensitivity variability is concerned, the results on NGC 188 indicate that these are not more than $\sim$5\%.

\section{Data Availability}

The data and codes that support the findings of this study are available in the repository \verb|UVIT-sensitivity-analysis| at \url{https://github.com/akanksha-dagore/UVIT-sensitivity-analysis} \citep{dagore_2025_18076974}.

\section{Acknowledgements}

The UVIT project is a part of the \textit{AstroSat} mission by the Indian Space Research Organisation (ISRO) and includes collaboration from the Indian Institute of Astrophysics (IIA), Bengaluru, the Indian-University Centre for Astronomy and Astrophysics (IUCAA), Pune, Tata Institute of Fundamental Research (TIFR), Mumbai, many centres of ISRO, and the Canadian Space Agency (CSA). This publication uses the data from the \textit{AstroSat} mission of the Indian Space Research Organisation (ISRO), archived at the Indian Space Science Data Centre (ISSDC). This work has made use of data from the European Space Agency (ESA) mission {\it Gaia} (\url{https://www.cosmos.esa.int/gaia}), processed by the {\it Gaia} Data Processing and Analysis Consortium (DPAC, \url{https://www.cosmos.esa.int/web/gaia/dpac/consortium}). Funding for the DPAC has been provided by national institutions, in particular the institutions participating in the {\it Gaia} Multilateral Agreement. This research has made use of the SIMBAD database, operated at CDS, Strasbourg, France.

\vspace{5mm}
\facilities{UVIT, \textit{Gaia}, SIMBAD}

\software{\textsc{Aafitrans} \citep{prajwel_aafitrans, BEROIZ2020100384}; \textsc{Astrometry.net} \citep{lang2010astrometry}; \textsc{Astropy} \citep{astropy:2013, astropy:2018, astropy:2022}; \textsc{Curvit} \citep{joseph2021curvit}; \textsc{Jupyter} \citep{Kluyver2016jupyter}; \textsc{Matplotlib} \citep{Hunter:2007}; \textsc{Numpy} \citep{harris2020array}; \textsc{Pandas} \citep{reback2020pandas}; \textsc{Photutils} \citep{larry_bradley_2022_6825092};  \textsc{SaoImageDS9} \citep{2003ASPC..295..489J}; \textsc{Scipy} \citep{2020SciPy-NMeth}; \textsc{Spyder} \citep{raybaut2009spyder}; \textsc{Topcat} \citep{2005ASPC..347...29T}
}

%



\clearpage

\appendix

\section{Image binning and rescaling}  
\label{appendix: binning_rescaling}

Here, we define the image binning and rescaling operations used in appendices \ref{appendix: background_estimation} and \ref{ssec: source_detection}. Let $A_{xy}$ denote a given pixel in the $x$th column and $y$th row in the original image. If the original image is binned using binning factors $m$ and $n$ pixels along the horizontal and vertical directions, respectively, then the binned pixel $B_{ij}$ in the $i$th column and $j$th row in the binned image can be represented by the equation,
\begin{equation}
    B_{ij} = \frac{1}{m\times n} \sum_{x=(i-1).m + 1}^{i.m} \ \sum_{y=(j-1).n+1}^{j.n}A_{xy} \ .  \label{eq:binning}
\end{equation}

The binned image constructed using equation (\ref{eq:binning}) can be rescaled back to the original dimension by repeating the binned pixel $m$ and $n$ times along the horizontal and vertical directions, respectively. Let $B_{xy}$ denote a given pixel in the $x$th column and $y$th row in the binned image. The rescaled pixel $R_{ij}$ in the $i$th column and $j$th row in the rescaled image can be represented by the equation, 
\begin{equation}
    R_{ij} = B_{xy} \,  
    \label{eq:rescaling}
\end{equation}
\begin{align}
    \mathrm{where,} \ x & = \left\lfloor \frac{i-1}{m} \right\rfloor + 1 \ , \\
     y & = \left\lfloor \frac{j-1}{n} \right\rfloor + 1 \ ,
\end{align}
with $m$ and $n$ being the rescaling factors. 

In this study, we have kept $m=n$ during binning and rescaling operations using equations (\ref{eq:binning}) and (\ref{eq:rescaling}). Additionally, the rescaling factors were kept equal to the binning factors.

\section{Background map generation} 
\label{appendix: background_estimation}

The UVIT images have relatively low background noise levels compared to the visible or infrared images, but we still need to correct the background to obtain accurate photometry. The background may vary in time and across the field of view. Only large-scale and smooth background variations are expected across the field of view in each image. While estimating the background levels from a visually selected nearby background region for each source in every episode image is possible, it is an inefficient and involving method while working with multiple sources and images.

Therefore, we developed a method to generate 2D background maps for the episode images. To create a background map, the common method adopted for optical imaging data involves placing an appropriately sized grid on the input image, taking the mean, median, or another statistic of each grid box to get the local background levels, and then interpolating the resulting low-resolution background map to match the input image resolution, with some form of filtering to suppress the under- or overestimation of the background levels. We have adopted this method with the median as the background estimator to generate the background maps, but with some modifications to account for the Poisson distribution of the background count levels of the UVIT images. A description of our background map generation method is given below.

The UVIT background count distribution is Poisson, which can be represented by the following equation:
\begin{equation}
    P(X\!=\!k) = \frac{\lambda^{k}e^{-\lambda}}{k!}, \ 
\end{equation}
where $X$ is a discrete random variable that represents the background counts per pixel, $\lambda$ is the expected value, and $P(X\!=\!k)$ gives the probability of $X$ being equal to $k$. Although $\lambda$ can be obtained by taking the mean of the distribution, the mean is highly influenced by outliers present in the real data. Therefore, we want to use the median to estimate $\lambda$ as it is resistant to outliers. However, the median cannot be used to estimate $\lambda$ when $\lambda$ is small (for example, the median might be almost always zero when $\lambda$ = 0.5). For larger values of $\lambda$ ($\lambda\!\geq\!10 $), the shape of the Poisson distribution resembles that of the normal form
and can be approximated to be a Gaussian distribution. Such a Gaussian approximation of the Poisson distribution is useful because both the median and mean are equal to the expected value for a Gaussian distribution; therefore, we can use the median to estimate $\lambda$ when this criterion is met.

To understand the uncertainties involved in using the median to estimate $\lambda$, we use the sharp bounds of the Poisson distribution median given by \cite{choi1994medians}:
\begin{equation}
    \lambda - ln(2) \le \nu <  \lambda + \frac{1}{3}, \ 
\end{equation}
where $\nu$ is the median of the Poisson distribution. The above equation suggests that taking the median as the Poisson expected value $\lambda$ will have an absolute uncertainty of $\sim$100/$\lambda$\%. For example, when $\lambda$ = 10, the median will be close to $\lambda$ with a possible $\sim$10\% error. The uncertainty becomes $\sim$5\% when $\lambda$ = 20. As $\lambda$ becomes large, the error reduces, and the median becomes very close to $\lambda$.

Consequently, we need $\lambda \approx 10$ to use the median approximation with a $\sim$10\% or better uncertainty. However, $\lambda$ $<$ 10 for all our episode images; the average photon count rate is $\sim$$5 \times 10^{-5}$ counts per second per sub-pixel for the NGC 188 F148W filter images, and $\lambda$ remains low even for those images with $\sim$2000 seconds exposure time ($\lambda$ = background countrate $\times$ exposure time). To overcome this problem, we made $\lambda \geq 10$ by binning the image sub-pixels using equation (\ref{eq:binning}) along the $x$ and $y$ directions to generate a `super-pixel’. Binning of the sub-pixels increases the background countrate per super-pixel. The binning factor $b$ was computed such that $\lambda \geq 10$ using the equation:
\begin{equation}
    {b} \geq \sqrt{\frac{10}{t_{exp} \times \mathrm{CPS}_{bkg}}} \ , 
    \label{eq:b_estimate}
\end{equation}
where $\mathrm{CPS}_{bkg}$ is the approximate background countrate in CPS and $t_{exp}$ is the exposure time in seconds. Since we have images with exposure times as low as $\sim$100 seconds, we used $t_{exp} = 100$ and $\mathrm{CPS}_{bkg} = 5 \times 10^{-5}$ in equation (\ref{eq:b_estimate}) and obtained $b\approx45$. We adopted a larger binning factor of 64 because we want to use binning factors that are powers of two for ease of dealing with the UVIT imaging data.

The UVIT FITS images are in the CPS unit and have a size of 4800$\times$4800 sub-pixels. We binned all episode images with a binning factor of 64 using equation (\ref{eq:binning}), resulting in binned images with sizes of 75$\times$75 super-pixels. Now, the background count distribution across the super-pixels corresponds to $\lambda \geq 10$ even for an exposure time as low as 100 seconds. The binned images were rescaled to the 4800$\times$4800 dimension using equation (\ref{eq:rescaling}) so that the background maps derived from the rescaled images would match the size of the input images. Background maps were generated from each rescaled image using the \verb|Background2D| function from the \verb|Photutils| package \citep{larry_bradley_2022_6825092}. We used a grid box size of 384$\times$384 sub-pixels (6$\times$6 super-pixels) and found the median background level in each grid box. 

\begin{figure}
\centering
\begin{subfigure}[b]{0.49\textwidth}
  \centering
  \includegraphics[width=\textwidth]{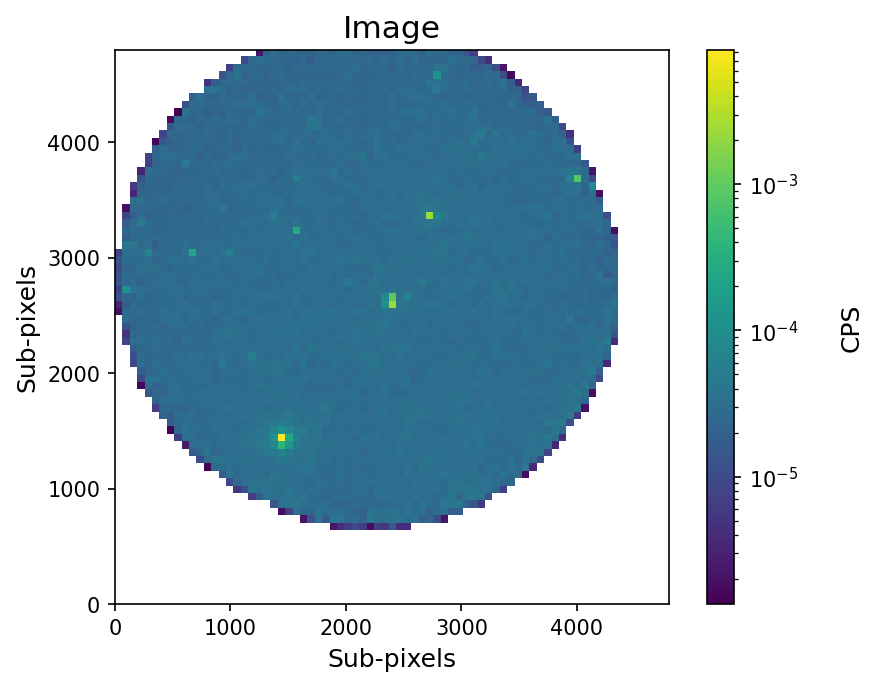}
\end{subfigure}
\begin{subfigure}[b]{0.49\textwidth}
  \centering
  \includegraphics[width=\textwidth]{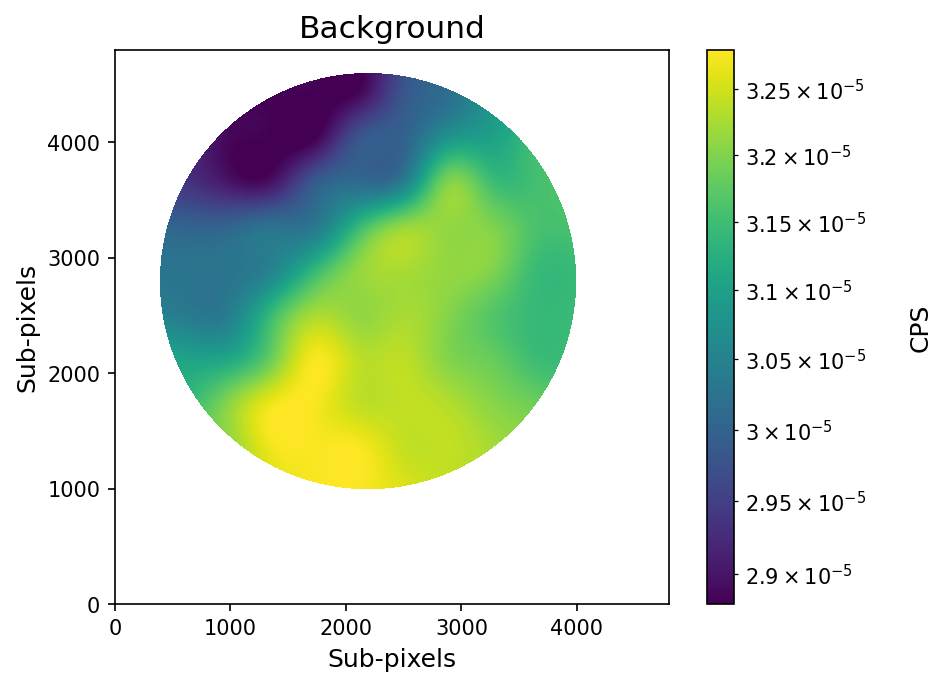}
\end{subfigure}
\caption{The binned and rescaled image (\pj{top} panel) and the background map (\pj{bottom} panel).}
\label{background_estimation}
\end{figure}

We restricted the background maps to a radius of 1800 sub-pixels from the centroid of the images for the NGC 188 data. The \pj{top} and \pj{bottom} panels in Fig. \ref{background_estimation} show a sample rescaled NGC 188 image and corresponding background map, respectively. The spread in the background levels (maximum $-$ minimum background) across the field of view is $\sim$$0.5\times10^{-5}$ CPS for all NGC 188 background maps. In the case of HZ 4, the background maps were restricted to boxes of 1500$\times$1500 sub-pixels centred on the image centroids. We found a background spread of $\sim$$0.2\times10^{-5}$ CPS in HZ 4. Fig. \ref{bkg} shows the background variations across NGC 188 and HZ 4 episode images. HZ4 UVIT images have higher FUV background levels than most NGC 188 images (this is also true for GALEX FUV imaging of these sources). The relatively high background levels observed in some episodes of NGC 188 data are discussed in Appendix \ref{high_bkg_sec}.

A Poisson treatment of the UVIT background can also be found in \cite{ananthamoorthy2024detection}.

\section{Source Detection} \label{ssec: source_detection}

We detected sources in the UVIT episode images using the DAOFIND algorithm implemented in the \verb|Photutils| package \citep{stetson1987daophot, larry_bradley_2022_6825092}. To improve source detectability, the episode images were pre-processed as follows. 

Each episode image in $4800\times4800$ sub-pixels was binned by a factor of 2 using equation (\ref{eq:binning}) to create an image in $2400\times2400$ binned-pixels. The binned image was convolved with a 2D Gaussian kernel having a standard deviation of 1.5 binned-pixels, resulting in a smoothed binned image with reduced noise levels. The source detectability is improved due to the increased signal-to-noise (SNR) ratio per binned-pixel, with a slight reduction in the detected source centroid accuracy. The smoothed binned image was rescaled back to the original $4800\times4800$ dimension following equation (\ref{eq:rescaling}) using a rescaling factor of 2.

DAOFIND was run on the pre-processed images to detect sources. The DAOFIND source detection threshold was $\kappa\times B$, where $B$ is the median value of the background map, and $\kappa$ is a constant. $\kappa$ was varied in the NGC 188 images such that 20-200 source detections were made, while $\kappa$ = 40 was used for all HZ 4 images. 
To avoid mistaking the noisy pixels at the edges as sources, we confined our detections in the NGC 188 images to a circular region of 1800 pixels.
The centre of this circular region is determined by taking the centroid of the 2D background array, which was calculated by taking the weighted sum of all the \textit{x} and \textit{y} pixel coordinates. Since the centroid was estimated from the image background, large background variations can cause the centroid to shift towards the denser background regions. However, there were no such large background variations in our input images. In the case of HZ 4, the detections were restricted to boxes of 1500$\times$1500 sub-pixels centred on the image centroids.

\section{Flux measurements and Uncertainties}

The corrected flux measurements and their associated uncertainties for HZ 4 and the three NGC 188 candidates used in Section \ref{sec:sens_analysis} are presented here. 
Tables \ref{table:flux_err_HZ4} and \ref{table:flux_err_NGC188} list the measurements for HZ 4 and the NGC 188 candidates, respectively.

\startlongtable
\begin{deluxetable}{ccccc}
\tablecaption{Corrected flux measurements and associated uncertainties for HZ 4 as shown in Fig. \ref{HZ4_combined}.
\label{table:flux_err_HZ4}}
\tablehead{\colhead{No.} & 
\colhead{Observation ID} & \colhead{Time} & \colhead{Source flux} & \colhead{Error} \\
\colhead{} & \colhead{} & \colhead{(MJD)} & \multicolumn{2}{c}{(CPS)}
}
\startdata
\multirow{3}{*}{1.} & \multirow{3}{*}{C02\textunderscore002T01\textunderscore9000000888} & 57738.420 & 23.385 & 0.259 \\
& & 57738.424 & 23.756 & 0.261 \\
& & 57738.428 & 24.361 & 0.264 \\ \hline
\multirow{4}{*}{2.} & \multirow{4}{*}{C03\textunderscore013T01\textunderscore9000001586} & 58032.171 & 23.606 & 0.275 \\
& & 58032.239 & 23.874 & 0.261 \\
& & 58032.242 & 23.873 & 0.313 \\
& & 58032.306 & 23.448 & 0.321 \\ \hline
\multirow{4}{*}{3.} & \multirow{4}{*}{C04\textunderscore010T01\textunderscore9000003158} & 58738.384 & 23.847 & 0.267 \\
& & 58738.389 & 22.915 & 0.262 \\
& & 58738.393 & 23.481 & 0.305 \\
& & 58738.451 & 23.330 & 0.390 \\ \hline
\multirow{2}{*}{4.} & \multirow{2}{*}{C02\textunderscore002T01\textunderscore9000006092} & 60364.549 & 23.574 & 0.265 \\
& & 60364.554 & 23.481 & 0.265 
\enddata
\end{deluxetable}


\startlongtable
\begin{deluxetable}{ccccccccc}
\tablecaption{Corrected flux measurements and associated uncertainties for the three NGC 188 candidates as shown in Fig. \ref{line_fit}.
\label{table:flux_err_NGC188}}
\tablehead{\colhead{No.} & 
\colhead{Observation ID} & \colhead{Time} & \multicolumn{2}{c}{Candidate 1} & \multicolumn{2}{c}{Candidate 2} & \multicolumn{2}{c}{Candidate 3} \\
\colhead{} & \colhead{} & \colhead{(MJD)} & \multicolumn{2}{c}{(CPS)} & \multicolumn{2}{c}{(CPS)} & \multicolumn{2}{c}{(CPS)} \\
\cline{4-9}
\colhead{} & \colhead{} & \colhead{} & \colhead{Source flux} & \colhead{Error} & \colhead{Source flux} & \colhead{Error} & \colhead{Source flux} & \colhead{Error}
}
\startdata
\multirow{2}{*}{1.} & \multirow{2}{*}{T01\textunderscore034T01\textunderscore9000000392} & 57469.571 & 16.918 & 0.171 & 11.044 & 0.140 & 0.864 & 0.042 \\
& & 57469.625 & 17.676 & 0.192 & 11.161 & 0.154 & 0.888 & 0.047 \\ \hline
2. & G05\textunderscore209T01\textunderscore9000000458 & 57526.814 & 17.132 & 0.123 & 11.250 & 0.101 & 0.896 & 0.031 \\ \hline
3. & G05\textunderscore258T01\textunderscore9000000666 & 57646.251 & 17.391 & 0.165 & 11.072 & 0.133 & 0.942 & 0.042 \\ \hline
4. & C02\textunderscore016T01\textunderscore9000000992 & 57783.494 & 17.281 & 0.118 & 11.437 & 0.098 & 0.995 & 0.031 \\ \hline
\multirow{3}{*}{5.} & \multirow{3}{*}{C02\textunderscore030T01\textunderscore9000001168} & 57859.093 & 17.401 & 0.234 & 11.395 & 0.192 & 0.899 & 0.059 \\
& & 57859.214 & 17.405 & 0.197 & 11.305 & 0.161 & 0.893 & 0.049 \\
& & 57859.284 & 16.637 & 0.214 & 11.155 & 0.178 & 0.837 & 0.053 \\ \hline
6. & C03\textunderscore015T01\textunderscore9000001788 & 58108.584 & 16.984 & 0.333 & 10.929 & 0.271 & 0.869 & 0.083 \\ \hline
\multirow{2}{*}{7.} & \multirow{2}{*}{T02\textunderscore002T01\textunderscore9000001914} & 58172.451 & 17.349 & 0.095 & 11.149 & 0.077 & 0.947 & 0.024 \\
& & 58172.514 & 17.227 & 0.127 & 11.099 & 0.103 & 0.914 & 0.032 \\ \hline
8. & C03\textunderscore015T02\textunderscore9000002008 & 58212.229 & 17.203 & 0.197 & 11.061 & 0.160 & 0.958 & 0.051 \\ \hline
9. & C03\textunderscore015T04\textunderscore9000002240 & 58319.753 & 17.361 & 0.126 & 11.009 & 0.102 & 0.937 & 0.032 \\ \hline
\multirow{3}{*}{10.} & \multirow{3}{*}{C03\textunderscore015T04\textunderscore9000002328} & 58356.101 & 16.804 & 0.253 & 10.707 & 0.205 & 0.965 & 0.066 \\
& & 58356.218 & 17.270 & 0.223 & 11.168 & 0.182 & 0.955 & 0.058 \\
& & 58356.286 & 17.472 & 0.177 & 11.134 & 0.143 & 0.871 & 0.043 \\ \hline
\multirow{3}{*}{11.} & \multirow{3}{*}{C03\textunderscore015T03\textunderscore9000002366} & 58375.045 & 17.445 & 0.175 & 10.634 & 0.139 & 0.898 & 0.043 \\
& & 58375.096 & 17.558 & 0.200 & 11.204 & 0.162 & 0.889 & 0.049 \\
& & 58375.115 & 17.528 & 0.331 & 10.473 & 0.261 & 0.812 & 0.079 \\ \hline
12. & C03\textunderscore015T04\textunderscore9000002382 & 58382.886 & 17.197 & 0.121 & 11.229 & 0.099 & 0.894 & 0.030 \\ \hline
13. & C04\textunderscore009T01\textunderscore9000002448 & 58414.413 & 17.227 & 0.121 & 10.992 & 0.098 & 0.942 & 0.031 \\ \hline
14. & C04\textunderscore009T02\textunderscore9000002618 & 58487.772 & 17.469 & 0.122 & 11.453 & 0.100 & 0.967 & 0.031 \\ \hline
15. & C05\textunderscore013T02\textunderscore9000003364 & 58830.612 & 17.365 & 0.121 & 11.083 & 0.098 & 0.940 & 0.031 \\ \hline
16. & C05\textunderscore013T03\textunderscore9000003376 & 58836.904 & 17.505 & 0.183 & 10.701 & 0.145 & 0.875 & 0.045 \\ \hline
17. & C05\textunderscore013T04\textunderscore9000003660 & 58981.357 & 17.120 & 0.120 & 11.174 & 0.099 & 0.939 & 0.031 \\ \hline
\multirow{2}{*}{18.} & \multirow{2}{*}{T03\textunderscore247T01\textunderscore9000003914} & 59127.105 & 17.363 & 0.295 & 11.086 & 0.240 & 0.943 & 0.076 \\
& & 59127.155 & 17.730 & 0.177 & 11.331 & 0.144 & 0.856 & 0.043 \\ \hline
19. & C05\textunderscore013T06\textunderscore9000003920 & 59137.921 & 17.053 & 0.150 & 10.965 & 0.122 & 0.876 & 0.037 \\ \hline
20. & C06\textunderscore010T01\textunderscore9000004504 & 59398.264 & 16.841 & 0.155 & 10.623 & 0.125 & 0.879 & 0.039 \\ \hline
\multirow{2}{*}{21.} & \multirow{2}{*}{C06\textunderscore010T02\textunderscore9000004618} & 59428.848 & 16.879 & 0.105 & 10.699 & 0.085 & 0.884 & 0.026 \\
& & 59428.915 & 16.654 & 0.136 & 10.613 & 0.110 & 0.876 & 0.034 \\ \hline
\multirow{2}{*}{22.} & \multirow{2}{*}{C06\textunderscore010T03\textunderscore9000004672} & 59453.406 & 16.707 & 0.128 & 10.899 & 0.105 & 0.851 & 0.032 \\
& & 59453.472 & 16.839 & 0.105 & 10.785 & 0.085 & 0.870 & 0.026 \\ \hline
\multirow{2}{*}{23.} & \multirow{2}{*}{C06\textunderscore010T05\textunderscore9000004736} & 59512.332 & 16.711 & 0.102 & 11.159 & 0.084 & 0.929 & 0.026 \\
& & 59512.392 & 16.837 & 0.283 & 11.185 & 0.233 & 0.965 & 0.074 \\ \hline
24. & T04\textunderscore065T01\textunderscore9000004750 & 59519.236 & 16.528 & 0.313 & 11.743 & 0.267 & 0.906 & 0.080 \\ \hline
\multirow{3}{*}{25.} & \multirow{3}{*}{C06\textunderscore010T05\textunderscore9000005424} & 59928.218 & 16.590 & 0.183 & 11.191 & 0.152 & 0.979 & 0.048 \\
& & 59928.284 & 16.451 & 0.110 & 10.905 & 0.091 & 0.922 & 0.028 \\
& & 59928.350 & 16.584 & 0.128 & 11.029 & 0.106 & 0.941 & 0.033 \\ \hline
\multirow{2}{*}{26.} & \multirow{2}{*}{C06\textunderscore010T05\textunderscore9000005436} & 59936.404 & 16.347 & 0.123 & 10.845 & 0.102 & 0.890 & 0.031 \\
& & 59936.470 & 16.871 & 0.097 & 10.726 & 0.079 & 0.917 & 0.025 \\ \hline
\multirow{2}{*}{27.} & \multirow{2}{*}{C06\textunderscore010T05\textunderscore9000005484} & 59959.873 & 17.005 & 0.107 & 11.026 & 0.088 & 0.909 & 0.027 \\
& & 59959.942 & 17.019 & 0.119 & 11.073 & 0.097 & 0.939 & 0.031 \\ \hline
\multirow{2}{*}{28.} & \multirow{2}{*}{T05\textunderscore101T01\textunderscore9000005578} & 60056.976 & 17.022 & 0.144 & 11.030 & 0.118 & 0.946 & 0.037 \\
& & 60057.043 & 17.278 & 0.281 & 10.812 & 0.226 & 1.055 & 0.076 \\ \hline
\multirow{2}{*}{29.} & \multirow{2}{*}{T05\textunderscore101T01\textunderscore9000005586} & 60062.993 & 17.361 & 0.141 & 10.681 & 0.112 & 0.895 & 0.035 \\
& & 60063.062 & 16.668 & 0.298 & 10.662 & 0.242 & 0.889 & 0.076 \\ \hline
\multirow{2}{*}{30.} & \multirow{2}{*}{C06\textunderscore010T05\textunderscore9000005670} & 60094.024 & 16.676 & 0.176 & 10.553 & 0.142 & 0.943 & 0.046 \\
& & 60094.215 & 16.507 & 0.134 & 10.408 & 0.109 & 0.867 & 0.035 \\ \hline
\multirow{2}{*}{31.} & \multirow{2}{*}{C06\textunderscore010T01\textunderscore9000005800} & 60154.930 & 16.300 & 0.112 & 10.018 & 0.089 & 0.838 & 0.028 \\
& & 60154.996 & 16.870 & 0.268 & 10.077 & 0.211 & 0.901 & 0.068 \\ \hline
\multirow{2}{*}{32.} & \multirow{2}{*}{C06\textunderscore010T01\textunderscore9000005816} & 60169.397 & 15.978 & 0.282 & 10.260 & 0.230 & 0.834 & 0.070 \\
& & 60169.451 & 16.119 & 0.201 & 10.615 & 0.166 & 0.764 & 0.050 \\ \hline
33. & C09\textunderscore011T02\textunderscore9000006310 & 60485.733 & 15.926 & 0.145 & 9.963 & 0.117 & 0.774 & 0.036 \\ \hline
\multirow{2}{*}{34.} & \multirow{2}{*}{C09\textunderscore011T03\textunderscore9000006410} & 60547.257 & 15.994 & 0.146 & 10.126 & 0.118 & 0.828 & 0.037 \\
& & 60547.323 & 15.872 & 0.106 & 10.190 & 0.086 & 0.876 & 0.028 \\ \hline
35. & C09\textunderscore011T01\textunderscore9000006418 & 60550.356 & 15.891 & 0.165 & 10.174 & 0.134 & 0.825 & 0.043
\enddata
\end{deluxetable}

\section{Building Astrometry.net index files for UVIT} \label{appendix: index_fils}

This appendix provides information on the custom-made index files used to perform astrometry on the UVIT images. We had limitations in using the index files provided by Astrometry.net\footnote{\url{http://data.astrometry.net/}}; the 5200 series index files, generated from visible channel catalogues failed to provide astrometric solutions in many of the UVIT images. The relatively lower spatial resolution of GALEX compared to UVIT affected the astrometric accuracy when 6000 and 6100 series index files were used. Therefore, new index files were generated using \verb|Astrometry.net| software \citep{lang2010astrometry} from a subset of the \textit{Gaia} data release 3 (DR3) catalogue \citep{collaboration2016description, 2023A&A...674A...1G}. This subset contains blue and bright sources that are expected to be visible in the UVIT images, and the derived index files are suitable for astrometry on UV images. The steps followed to build the index files are provided in the following subsections.

\subsection{Creating the subset catalogue}

\begin{figure}
\centering
\begin{subfigure}[b]{0.48\textwidth}
  \centering
  \includegraphics[width=\textwidth]{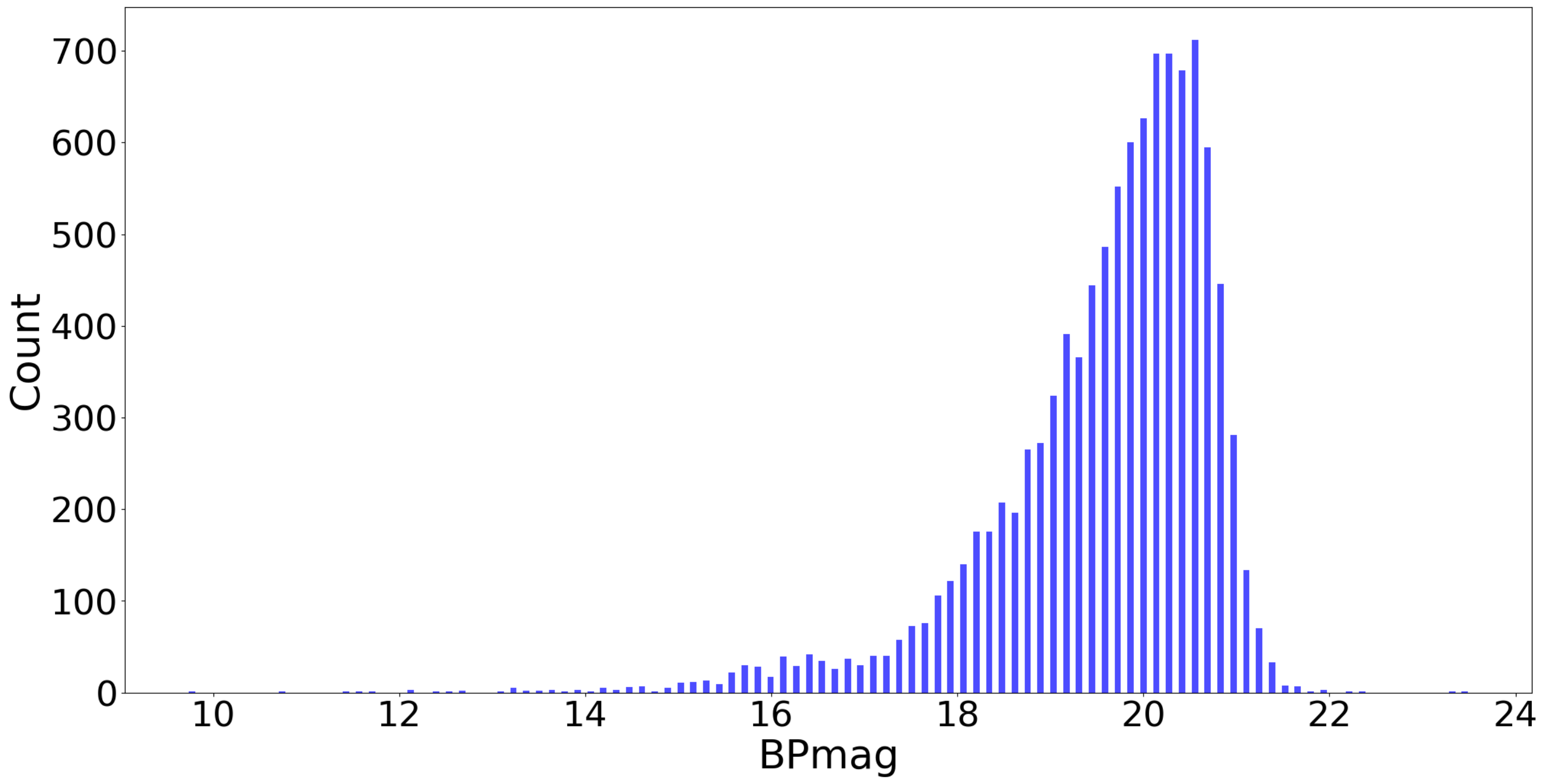}
  \caption{}
  \label{bpmag}
\end{subfigure}
\begin{subfigure}[b]{0.48\textwidth}
  \centering
  \includegraphics[width=\textwidth]{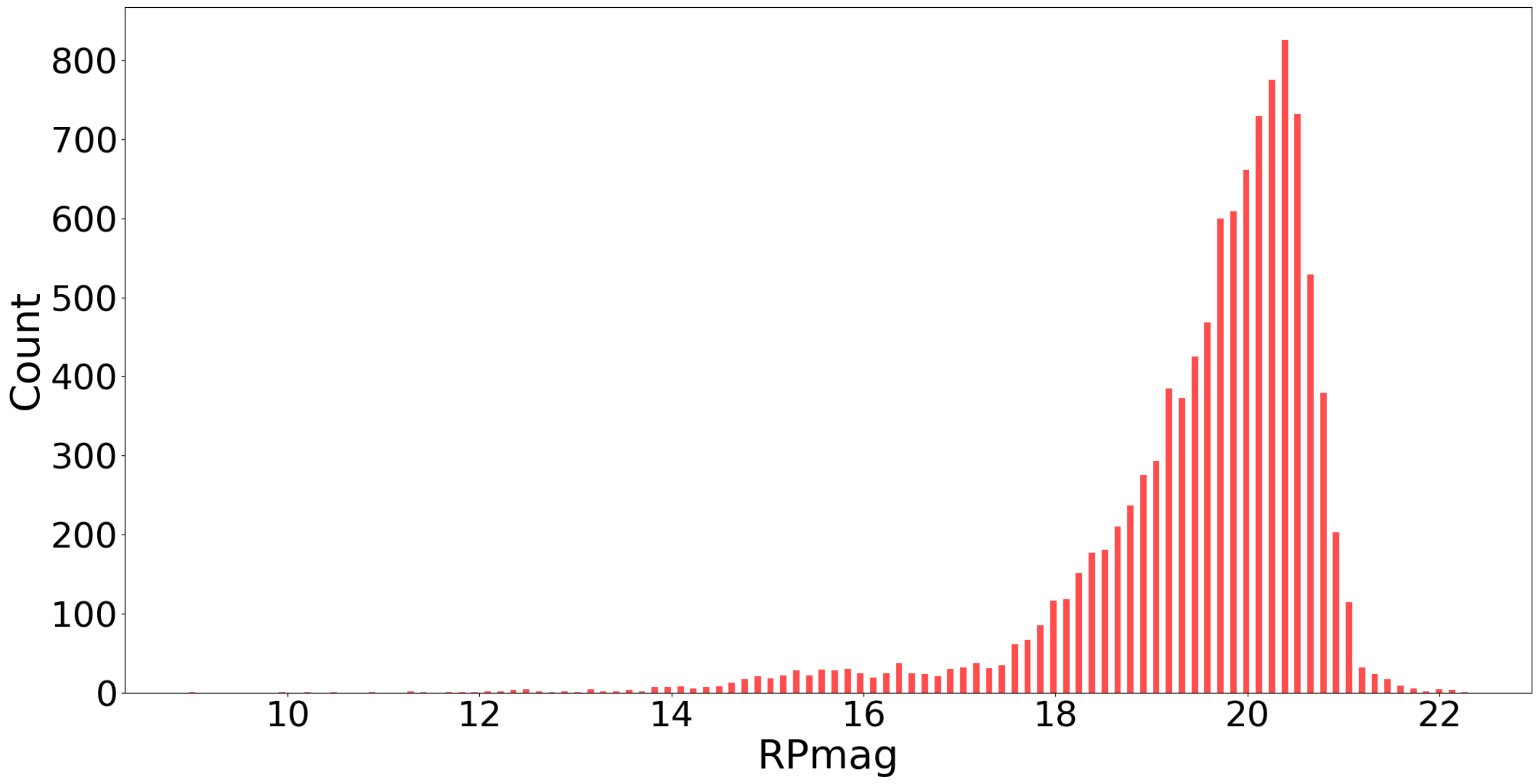}
  \caption{}
  \label{rpmag}
\end{subfigure}
\begin{subfigure}[b]{0.48\textwidth}
  \centering
  \includegraphics[width=\textwidth]{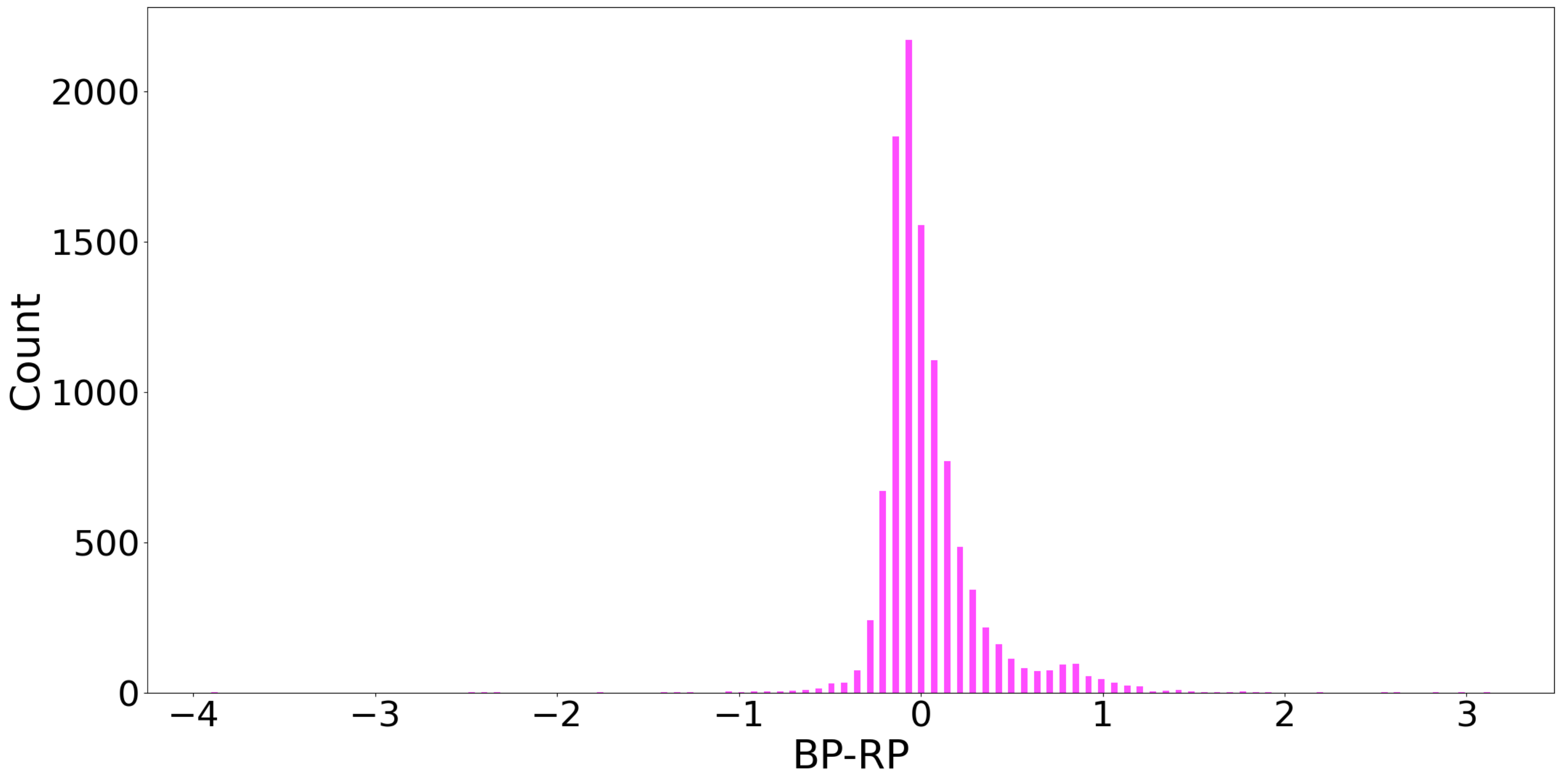}
  \caption{}
  \label{bp-rp}
\end{subfigure}
\caption{The three-panelled plot represents the distribution of the sources in the UVIT SMC catalogue cross-matched with the \textit{Gaia} catalogue in terms of BPmag, RPmag and BP-RP.}
\label{bp_rp_bp-rp}
\end{figure}

The first step in the index files generation process was to create a subset of the \textit{Gaia} DR3 catalogue (subset catalogue) that contains blue and bright sources. The source selection criterion to generate the subset catalogue was established using the UVIT Small Magellanic Cloud (SMC) catalogue that contains 11,241 point-sources \citep{devaraj2023uvit}. The celestial coordinates ($\alpha$, $\delta$) of the UVIT SMC catalogue sources were cross-matched with that of \textit{Gaia} DR3 catalogue, and it resulted in a matched list of 10,784 sources (96\% of the SMC catalogue sources). The histograms in Fig. \ref{bp_rp_bp-rp} depict the \textit{Gaia} blue passband magnitude (BPmag), red passband magnitude (RPmag), and BPmag $-$ RPmag colour (BP-RP) distribution of the UVIT SMC-\textit{Gaia} DR3 matched sources. From a visual inspection of the histograms and multiple trials, the following source selection criterion was finalised:
\begin{align}
    & \mathrm{( BPmag <= 12 ) \ OR} \nonumber \\ 
    & ( \mathrm{BP-RP <= 0.8 \ AND \ BPmag <= 21 )} .
\end{align}

Table \ref{table:bpmag_rp_mag_bp-rp} shows the percentage of sources falling under each part of the source selection criteria for the UVIT SMC-\textit{Gaia} DR3 matched catalogue. The whole \textit{Gaia} DR3 catalogue gave a list of 64,597,375 sources that matched our source-selection criteria. The data for these sources was downloaded for the J2000 reference epoch from one of \textit{Gaia}'s partner data centres, Centre de Données astronomiques de Strasbourg (CDS-France)\footnote{\url{https://cds.unistra.fr/}}. The subset catalogue was then sorted based on the BPmag column in ascending order and converted into the FITS table format.

\begin{deluxetable}{cc}
\tablecaption{Percentage of sources falling under each part of the source selection criteria for the UVIT SMC-\textit{Gaia} DR3 matched catalogue.
\label{table:bpmag_rp_mag_bp-rp}}
\tablehead{
\colhead{Criteria} & \colhead{Source}
}
\startdata
BPmag $\leq$ 12 & 0.046\%  \\
BPmag $\leq$ 21 & 94.807 \% \\
BP-RP $\leq$ 0.8 & 94.371\% \\
(BP-RP $\leq$ 0.8) AND (BPmag $\leq$ 21) & 91.747\% \\
\enddata
\end{deluxetable}

\subsection{Generating the index files}

The subset catalogue was divided into 48 FITS tables, corresponding to the 48 HEALPix equal-area subdivisions of the sky. This division was achieved using the \verb|hsplit| command from \verb|Astrometry.net|, which uses the $\alpha$ and $\delta$ column values of the subset catalogue to split the sky into equal area subdivisions according to the HEALPix specification \citep{2005ApJ...622..759G}. Subsequently, the \verb|build-astrometry-index| command was used to generate index files from each of the 48 FITS tables. During the index file generation, we selected \verb|build-astrometry-index| preset values from 1 to 9, which generates index files that contain features of different angular sizes. This selection led to the creation of 9 index files for each of the 48 HEALpix tiles, each corresponding to a different scale, resulting in 432 UVIT index files. Using these, we were able to obtain the astrometric solution for the UVIT images of NGC 188. The generated index files will be generally useful for obtaining astrometric solutions for UVIT images.

\section{Cases of high background observed in the UVIT images}
\label{high_bkg_sec}

\begin{figure}
    \centering
    \includegraphics[width=\columnwidth]{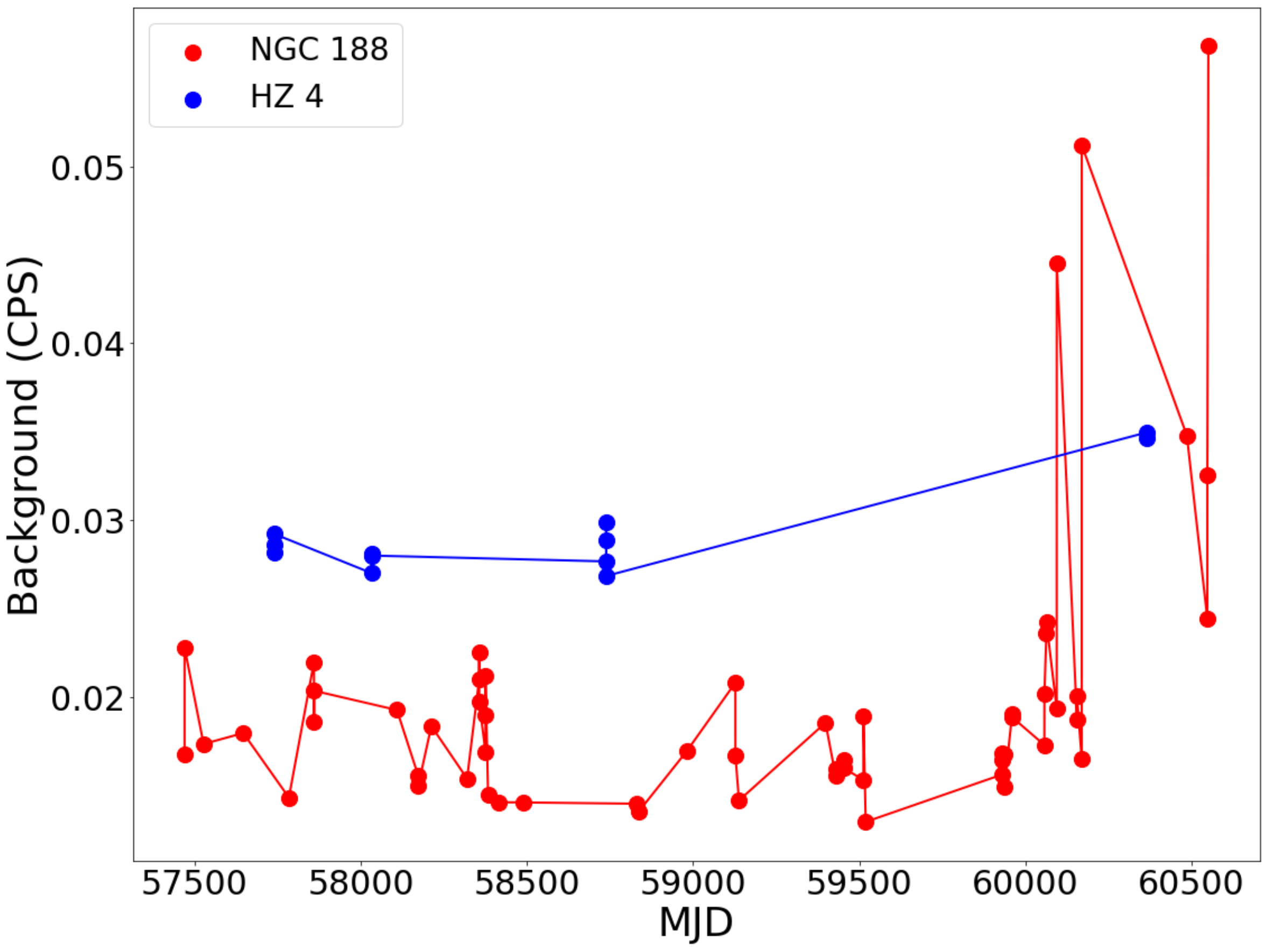}
    \caption{Median background levels corresponding to a 12 sub-pixel radius aperture in the F148W images of NGC 188 and HZ 4 across different episodes.}
    \label{bkg}
\end{figure}

This section presents our findings regarding relatively high background levels in recent NGC 188 episode images. For every input image, we estimated the background enclosed under a 12 sub-pixel radius aperture by multiplying the image background map median value, $B$, with a factor of $\pi\times12^{2}$. These background levels are shown in Fig. \ref{bkg} for HZ4 and NGC 188. 
Notably, the most recent NGC 188 episode images exhibit significant variations in background levels. 
HZ4 episode images have a higher background level than most NGC 188 episode images. GALEX imaging of HZ4 similarly shows elevated background levels relative to GALEX imaging of NGC 188.
Such sky-location dependent variations in UV background levels are known \citep{henry2014mystery}.


\subsection{Temporal variation of the background}

\begin{figure*}
    \centering
  \includegraphics[width=0.8\textwidth]{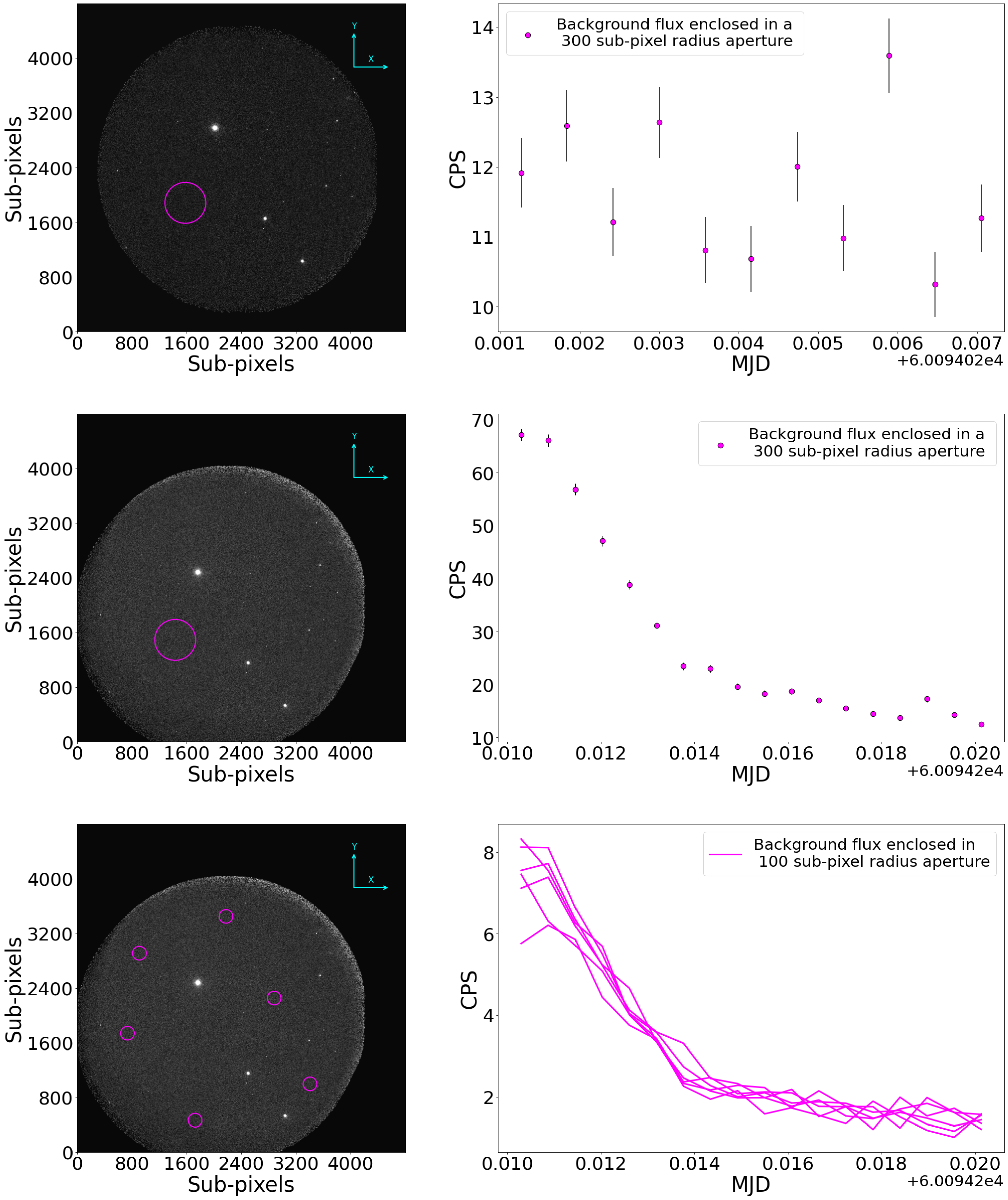}
    \caption{The three-panelled subplot shows the temporal variation of the background flux for the episodes from observation ID C06\_010T05\_9000005670. The middle and bottom panels display episodes with high background levels, while the top panel shows the preceding episode with normal background levels. On the left side of each panel, the UVIT F148W image in instrument coordinates highlights the selected background regions, marked by circular apertures. The corresponding light curves for these regions are shown on the right side of each panel.}
    \label{high_bkg_cases}
\end{figure*}

We checked whether the background exhibited spatial or temporal variations within the affected episodes. To study the temporal variations, we used \verb|Curvit| \citep{joseph2021curvit} to produce light curves for the background regions. The background regions were selected through visual inspection. We generated background light curves using a 300 sub-pixel radius aperture and a time bin of 50 seconds. Fig. \ref{high_bkg_cases} shows the background light curves for a high background episode and the preceding normal background episode in observation ID C06\_010T05\_9000005670.
In the figure, the top panel corresponds to the preceding episode with a normal background, while the middle panel depicts the high background episode. On the left side of the panel, the selected background region in the UVIT image field is highlighted by a 300 sub-pixel radius aperture, and the corresponding light curve is displayed on the right side. In the high background episode, the background varies, with the levels being high initially and then reducing over time. No such large background variation was observed in the normal background episode.
\pj{Temporal variations in background levels were found in all six episodes with relatively high background levels.}


To determine whether the elevated background levels were spatially localised, we placed six smaller apertures, each with a 100 sub-pixel radius, in randomly selected background regions of the high-background UVIT episodes. Light curves were generated for these regions, as shown in the bottom panel of Fig. \ref{high_bkg_cases}.
In high-background episodes, the background flux variations across the different regions were consistent. This consistency indicates that the elevated background levels are not confined to specific areas and are a widespread feature across the image. 

As per the \textit{AstroSat} Handbook (Version: 1.10)\footnote{\url{https://www.issdc.gov.in/astro.html}}, possible sources that can contribute to the UVIT background include the bright limb of the Earth, zodiacal light, and geocoronal lines during daytime observations. 
\pj{Our preliminary checks suggest that the background levels are varying depending on the position of the spacecraft in orbit.}
\pj{Notably, the rise in background levels coincides with the solar cycle 25 maximum.
A detailed analysis of background variations will be presented in a separate study.
}


\bibliography{references}{}

@article{astropy:2013,
Adsurl = {http://adsabs.harvard.edu/abs/2013A%26A...558A..33A},
Archiveprefix = {arXiv},
Author = {{Astropy Collaboration} and {Robitaille}, T.~P. and {Tollerud}, E.~J. and {Greenfield}, P. and {Droettboom}, M. and {Bray}, E. and {Aldcroft}, T. and {Davis}, M. and {Ginsburg}, A. and {Price-Whelan}, A.~M. and {Kerzendorf}, W.~E. and {Conley}, A. and {Crighton}, N. and {Barbary}, K. and {Muna}, D. and {Ferguson}, H. and {Grollier}, F. and {Parikh}, M.~M. and {Nair}, P.~H. and {Unther}, H.~M. and {Deil}, C. and {Woillez}, J. and {Conseil}, S. and {Kramer}, R. and {Turner}, J.~E.~H. and {Singer}, L. and {Fox}, R. and {Weaver}, B.~A. and {Zabalza}, V. and {Edwards}, Z.~I. and {Azalee Bostroem}, K. and {Burke}, D.~J. and {Casey}, A.~R. and {Crawford}, S.~M. and {Dencheva}, N. and {Ely}, J. and {Jenness}, T. and {Labrie}, K. and {Lim}, P.~L. and {Pierfederici}, F. and {Pontzen}, A. and {Ptak}, A. and {Refsdal}, B. and {Servillat}, M. and {Streicher}, O.},
Doi = {10.1051/0004-6361/201322068},
Eid = {A33},
Eprint = {1307.6212},
Journal = {\aap},
Month = oct,
Pages = {A33},
Primaryclass = {astro-ph.IM},
Title = {{Astropy: A community Python package for astronomy}},
Volume = 558,
Year = 2013}

@ARTICLE{astropy:2018,
       author = {{Astropy Collaboration} and {Price-Whelan}, A.~M. and
         {Sip{\H{o}}cz}, B.~M. and {G{\"u}nther}, H.~M. and {Lim}, P.~L. and
         {Crawford}, S.~M. and {Conseil}, S. and {Shupe}, D.~L. and
         {Craig}, M.~W. and {Dencheva}, N. and {Ginsburg}, A. and {Vand
        erPlas}, J.~T. and {Bradley}, L.~D. and {P{\'e}rez-Su{\'a}rez}, D. and
         {de Val-Borro}, M. and {Aldcroft}, T.~L. and {Cruz}, K.~L. and
         {Robitaille}, T.~P. and {Tollerud}, E.~J. and {Ardelean}, C. and
         {Babej}, T. and {Bach}, Y.~P. and {Bachetti}, M. and {Bakanov}, A.~V. and
         {Bamford}, S.~P. and {Barentsen}, G. and {Barmby}, P. and
         {Baumbach}, A. and {Berry}, K.~L. and {Biscani}, F. and {Boquien}, M. and
         {Bostroem}, K.~A. and {Bouma}, L.~G. and {Brammer}, G.~B. and
         {Bray}, E.~M. and {Breytenbach}, H. and {Buddelmeijer}, H. and
         {Burke}, D.~J. and {Calderone}, G. and {Cano Rodr{\'\i}guez}, J.~L. and
         {Cara}, M. and {Cardoso}, J.~V.~M. and {Cheedella}, S. and {Copin}, Y. and
         {Corrales}, L. and {Crichton}, D. and {D'Avella}, D. and {Deil}, C. and
         {Depagne}, {\'E}. and {Dietrich}, J.~P. and {Donath}, A. and
         {Droettboom}, M. and {Earl}, N. and {Erben}, T. and {Fabbro}, S. and
         {Ferreira}, L.~A. and {Finethy}, T. and {Fox}, R.~T. and
         {Garrison}, L.~H. and {Gibbons}, S.~L.~J. and {Goldstein}, D.~A. and
         {Gommers}, R. and {Greco}, J.~P. and {Greenfield}, P. and
         {Groener}, A.~M. and {Grollier}, F. and {Hagen}, A. and {Hirst}, P. and
         {Homeier}, D. and {Horton}, A.~J. and {Hosseinzadeh}, G. and {Hu}, L. and
         {Hunkeler}, J.~S. and {Ivezi{\'c}}, {\v{Z}}. and {Jain}, A. and
         {Jenness}, T. and {Kanarek}, G. and {Kendrew}, S. and {Kern}, N.~S. and
         {Kerzendorf}, W.~E. and {Khvalko}, A. and {King}, J. and {Kirkby}, D. and
         {Kulkarni}, A.~M. and {Kumar}, A. and {Lee}, A. and {Lenz}, D. and
         {Littlefair}, S.~P. and {Ma}, Z. and {Macleod}, D.~M. and
         {Mastropietro}, M. and {McCully}, C. and {Montagnac}, S. and
         {Morris}, B.~M. and {Mueller}, M. and {Mumford}, S.~J. and {Muna}, D. and
         {Murphy}, N.~A. and {Nelson}, S. and {Nguyen}, G.~H. and
         {Ninan}, J.~P. and {N{\"o}the}, M. and {Ogaz}, S. and {Oh}, S. and
         {Parejko}, J.~K. and {Parley}, N. and {Pascual}, S. and {Patil}, R. and
         {Patil}, A.~A. and {Plunkett}, A.~L. and {Prochaska}, J.~X. and
         {Rastogi}, T. and {Reddy Janga}, V. and {Sabater}, J. and
         {Sakurikar}, P. and {Seifert}, M. and {Sherbert}, L.~E. and
         {Sherwood-Taylor}, H. and {Shih}, A.~Y. and {Sick}, J. and
         {Silbiger}, M.~T. and {Singanamalla}, S. and {Singer}, L.~P. and
         {Sladen}, P.~H. and {Sooley}, K.~A. and {Sornarajah}, S. and
         {Streicher}, O. and {Teuben}, P. and {Thomas}, S.~W. and
         {Tremblay}, G.~R. and {Turner}, J.~E.~H. and {Terr{\'o}n}, V. and
         {van Kerkwijk}, M.~H. and {de la Vega}, A. and {Watkins}, L.~L. and
         {Weaver}, B.~A. and {Whitmore}, J.~B. and {Woillez}, J. and
         {Zabalza}, V. and {Astropy Contributors}},
        title = "{The Astropy Project: Building an Open-science Project and Status of the v2.0 Core Package}",
      journal = {\aj},
         year = 2018,
        month = sep,
       volume = {156},
       number = {3},
          eid = {123},
        pages = {123},
          doi = {10.3847/1538-3881/aabc4f},
archivePrefix = {arXiv},
       eprint = {1801.02634},
 primaryClass = {astro-ph.IM},
       adsurl = {https://ui.adsabs.harvard.edu/abs/2018AJ....156..123A}
}

@ARTICLE{astropy:2022,
       author = {{Astropy Collaboration} and {Price-Whelan}, Adrian M. and {Lim}, Pey Lian and {Earl}, Nicholas and {Starkman}, Nathaniel and {Bradley}, Larry and {Shupe}, David L. and {Patil}, Aarya A. and {Corrales}, Lia and {Brasseur}, C.~E. and {N{"o}the}, Maximilian and {Donath}, Axel and {Tollerud}, Erik and {Morris}, Brett M. and {Ginsburg}, Adam and {Vaher}, Eero and {Weaver}, Benjamin A. and {Tocknell}, James and {Jamieson}, William and {van Kerkwijk}, Marten H. and {Robitaille}, Thomas P. and {Merry}, Bruce and {Bachetti}, Matteo and {G{"u}nther}, H. Moritz and {Aldcroft}, Thomas L. and {Alvarado-Montes}, Jaime A. and {Archibald}, Anne M. and {B{'o}di}, Attila and {Bapat}, Shreyas and {Barentsen}, Geert and {Baz{'a}n}, Juanjo and {Biswas}, Manish and {Boquien}, M{'e}d{'e}ric and {Burke}, D.~J. and {Cara}, Daria and {Cara}, Mihai and {Conroy}, Kyle E. and {Conseil}, Simon and {Craig}, Matthew W. and {Cross}, Robert M. and {Cruz}, Kelle L. and {D'Eugenio}, Francesco and {Dencheva}, Nadia and {Devillepoix}, Hadrien A.~R. and {Dietrich}, J{"o}rg P. and {Eigenbrot}, Arthur Davis and {Erben}, Thomas and {Ferreira}, Leonardo and {Foreman-Mackey}, Daniel and {Fox}, Ryan and {Freij}, Nabil and {Garg}, Suyog and {Geda}, Robel and {Glattly}, Lauren and {Gondhalekar}, Yash and {Gordon}, Karl D. and {Grant}, David and {Greenfield}, Perry and {Groener}, Austen M. and {Guest}, Steve and {Gurovich}, Sebastian and {Handberg}, Rasmus and {Hart}, Akeem and {Hatfield-Dodds}, Zac and {Homeier}, Derek and {Hosseinzadeh}, Griffin and {Jenness}, Tim and {Jones}, Craig K. and {Joseph}, Prajwel and {Kalmbach}, J. Bryce and {Karamehmetoglu}, Emir and {Ka{l}uszy{'n}ski}, Miko{l}aj and {Kelley}, Michael S.~P. and {Kern}, Nicholas and {Kerzendorf}, Wolfgang E. and {Koch}, Eric W. and {Kulumani}, Shankar and {Lee}, Antony and {Ly}, Chun and {Ma}, Zhiyuan and {MacBride}, Conor and {Maljaars}, Jakob M. and {Muna}, Demitri and {Murphy}, N.~A. and {Norman}, Henrik and {O'Steen}, Richard and {Oman}, Kyle A. and {Pacifici}, Camilla and {Pascual}, Sergio and {Pascual-Granado}, J. and {Patil}, Rohit R. and {Perren}, Gabriel I. and {Pickering}, Timothy E. and {Rastogi}, Tanuj and {Roulston}, Benjamin R. and {Ryan}, Daniel F. and {Rykoff}, Eli S. and {Sabater}, Jose and {Sakurikar}, Parikshit and {Salgado}, Jes{'u}s and {Sanghi}, Aniket and {Saunders}, Nicholas and {Savchenko}, Volodymyr and {Schwardt}, Ludwig and {Seifert-Eckert}, Michael and {Shih}, Albert Y. and {Jain}, Anany Shrey and {Shukla}, Gyanendra and {Sick}, Jonathan and {Simpson}, Chris and {Singanamalla}, Sudheesh and {Singer}, Leo P. and {Singhal}, Jaladh and {Sinha}, Manodeep and {Sip{H{o}}cz}, Brigitta M. and {Spitler}, Lee R. and {Stansby}, David and {Streicher}, Ole and {{{S}}umak}, Jani and {Swinbank}, John D. and {Taranu}, Dan S. and {Tewary}, Nikita and {Tremblay}, Grant R. and {Val-Borro}, Miguel de and {Van Kooten}, Samuel J. and {Vasovi{'c}}, Zlatan and {Verma}, Shresth and {de Miranda Cardoso}, Jos{'e} Vin{'i}cius and {Williams}, Peter K.~G. and {Wilson}, Tom J. and {Winkel}, Benjamin and {Wood-Vasey}, W.~M. and {Xue}, Rui and {Yoachim}, Peter and {Zhang}, Chen and {Zonca}, Andrea and {Astropy Project Contributors}},
        title = "{The Astropy Project: Sustaining and Growing a Community-oriented Open-source Project and the Latest Major Release (v5.0) of the Core Package}",
      journal = {\apj},
         year = 2022,
        month = aug,
       volume = {935},
       number = {2},
          eid = {167},
        pages = {167},
          doi = {10.3847/1538-4357/ac7c74},
archivePrefix = {arXiv},
       eprint = {2206.14220},
 primaryClass = {astro-ph.IM},
       adsurl = {https://ui.adsabs.harvard.edu/abs/2022ApJ...935..167A}
}

@article{rani2021uocs,
       author = {{Rani}, Sharmila and {Subramaniam}, Annapurni and {Pandey}, Sindhu and {Sahu}, Snehalata and {Mondal}, Chayan and {Pandey}, Gajendra},
        title = "{UOCS. V. UV study of the old open cluster NGC 188 using AstroSat}",
      journal = {Journal of Astrophysics and Astronomy},
         year = 2021,
        month = oct,
       volume = {42},
       number = {2},
          eid = {47},
        pages = {47},
          doi = {10.1007/s12036-020-09683-2},
archivePrefix = {arXiv},
       eprint = {2012.00510},
 primaryClass = {astro-ph.SR},
       adsurl = {https://ui.adsabs.harvard.edu/abs/2021JApA...42...47R}
}

@article{subramaniam2016hot,
       author = {{Subramaniam}, Annapurni and {Sindhu}, N. and {Tandon}, S.~N. and {Kameswara Rao}, N. and {Postma}, J. and {C{\^o}t{\'e}}, Patrick and {Hutchings}, J.~B. and {Ghosh}, S.~K. and {George}, K. and {Girish}, V. and {Mohan}, R. and {Murthy}, J. and {Sankarasubramanian}, K. and {Stalin}, C.~S. and {Sutaria}, F. and {Mondal}, C. and {Sahu}, S.},
        title = "{A Hot Companion to a Blue Straggler in NGC 188 as Revealed by the Ultra-Violet Imaging Telescope (UVIT) on ASTROSAT}",
      journal = {The Astrophysical Journal},
         year = 2016,
        month = dec,
       volume = {833},
       number = {2},
          eid = {L27},
        pages = {L27},
          doi = {10.3847/2041-8213/833/2/L27},
       adsurl = {https://ui.adsabs.harvard.edu/abs/2016ApJ...833L..27S}
}

@article{henry2014mystery,
  title={The mystery of the cosmic diffuse ultraviolet background radiation},
  author={Henry, Richard Conn and Murthy, Jayant and Overduin, James and Tyler, Joshua},
  journal={The Astrophysical Journal},
  volume={798},
  number={1},
  pages={14},
  year={2014},
  publisher={IOP Publishing}
}

@article{ananthamoorthy2024detection,
  title={Detection of Faint Sources by the UltraViolet Imaging Telescope Onboard AstroSat Using Poisson Distribution of Background},
  author={Ananthamoorthy, B and Bhattacharya, Debbijoy and Sreekumar, P and Swathi, B},
  journal={The Astronomical Journal},
  volume={168},
  number={1},
  pages={22},
  year={2024},
  publisher={IOP Publishing}
}

@article{ghosh2022automated,
       author = {{Ghosh}, S.~K. and {Tandon}, S.~N. and {Singh}, S.~K. and {Shelat}, D.~S. and {Tahlani}, P. and {Singh}, A.~K. and {Srinivasan}, T.~P. and {Joseph}, P. and {Devaraj}, A. and {George}, K. and {Mohan}, R. and {Postma}, J. and {Stalin}, C.~S.},
        title = "{An automated pipeline for Ultra-Violet Imaging Telescope}",
      journal = {Journal of Astrophysics and Astronomy},
         year = 2022,
        month = dec,
       volume = {43},
       number = {2},
          eid = {77},
        pages = {77},
          doi = {10.1007/s12036-022-09842-7},
archivePrefix = {arXiv},
       eprint = {2203.07693},
 primaryClass = {astro-ph.IM},
       adsurl = {https://ui.adsabs.harvard.edu/abs/2022JApA...43...77G}
}

@article{tandon2020additional,
       author = {{Tandon}, S.~N. and {Postma}, J. and {Joseph}, P. and {Devaraj}, A. and {Subramaniam}, A. and {Barve}, I.~V. and {George}, K. and {Ghosh}, S.~K. and {Girish}, V. and {Hutchings}, J.~B. and {Kamath}, P.~U. and {Kathiravan}, S. and {Kumar}, A. and {Lancelot}, J.~P. and {Leahy}, D. and {Mahesh}, P.~K. and {Mohan}, R. and {Nagabhushana}, S. and {Pati}, A.~K. and {Rao}, N. Kameswara and {Sankarasubramanian}, K. and {Sriram}, S. and {Stalin}, C.~S.},
        title = "{Additional Calibration of the Ultraviolet Imaging Telescope on Board AstroSat}",
      journal = {\aj},
         year = 2020,
        month = apr,
       volume = {159},
       number = {4},
          eid = {158},
        pages = {158},
          doi = {10.3847/1538-3881/ab72a3},
archivePrefix = {arXiv},
       eprint = {2002.01159},
 primaryClass = {astro-ph.IM},
       adsurl = {https://ui.adsabs.harvard.edu/abs/2020AJ....159..158T}
}

@article{ghosh2021performance,
       author = {{Ghosh}, S.~K. and {Tandon}, S.~N. and {Joseph}, P. and {Devaraj}, A. and {Shelat}, D.~S. and {Stalin}, C.~S.},
        title = "{Performance of the UVIT Level-2 pipeline}",
      journal = {Journal of Astrophysics and Astronomy},
         year = 2021,
        month = oct,
       volume = {42},
       number = {2},
          eid = {29},
        pages = {29},
          doi = {10.1007/s12036-020-09686-z},
archivePrefix = {arXiv},
       eprint = {2012.13534},
 primaryClass = {astro-ph.IM},
       adsurl = {https://ui.adsabs.harvard.edu/abs/2021JApA...42...29G}
}

@article{stetson1987daophot,
  title={DAOPHOT: A computer program for crowded-field stellar photometry},
  author={Stetson, Peter B},
  journal={Publications of the Astronomical Society of the Pacific},
  volume={99},
  number={613},
  pages={191},
  year={1987},
  publisher={IOP Publishing}
}

@article{devaraj2023uvit,
       author = {{Devaraj}, A. and {Joseph}, P. and {Stalin}, C.~S. and {Tandon}, S.~N. and {Ghosh}, S.~K.},
        title = "{UVIT Observations of the Small Magellanic Cloud: Point-source Catalog}",
      journal = {\apj},
         year = 2023,
        month = apr,
       volume = {946},
       number = {2},
          eid = {65},
        pages = {65},
          doi = {10.3847/1538-4357/acba9c},
archivePrefix = {arXiv},
       eprint = {2302.01515},
 primaryClass = {astro-ph.GA},
       adsurl = {https://ui.adsabs.harvard.edu/abs/2023ApJ...946...65D}
}

@software{larry_bradley_2022_6825092,
author       = {Larry Bradley and
                Brigitta Sipőcz and
                Thomas Robitaille and
                Erik Tollerud and
                Zé Vinícius and
                Christoph Deil and
                Kyle Barbary and
                Tom J Wilson and
                Ivo Busko and
                Axel Donath and
                Hans Moritz Günther and
                Mihai Cara and
                P. L. Lim and
                Sebastian Meßlinger and
                Simon Conseil and
                Azalee Bostroem and
                Michael Droettboom and
                E. M. Bray and
                Lars Andersen Bratholm and
                Geert Barentsen and
                Matt Craig and
                Shivangee Rathi and
                Sergio Pascual and
                Gabriel Perren and
                Iskren Y. Georgiev and
                Miguel de Val-Borro and
                Wolfgang Kerzendorf and
                Yoonsoo P. Bach and
                Bruno Quint and
                Harrison Souchereau},
title        = {astropy/photutils: 1.5.0},
month        = jul,
year         = 2022,
publisher    = {Zenodo},
version      = {1.5.0},
doi          = {10.5281/zenodo.6825092},
url          = {https://doi.org/10.5281/zenodo.6825092}
}

@article{BEROIZ2020100384,
title = "Astroalign: A Python module for astronomical image registration",
journal = "Astronomy and Computing",
volume = "32",
pages = "100384",
year = "2020",
issn = "2213-1337",
doi = "https://doi.org/10.1016/j.ascom.2020.100384",
url = "http://www.sciencedirect.com/science/article/pii/S221313372030038X",
author = "M. Beroiz and J.B. Cabral and B. Sanchez"
}

@article{joseph2021curvit,
       author = {{Joseph}, P. and {Stalin}, C.~S. and {Tandon}, S.~N. and {Ghosh}, S.~K.},
        title = "{Curvit: An open-source Python package to generate light curves from UVIT data}",
      journal = {Journal of Astrophysics and Astronomy},
         year = 2021,
        month = oct,
       volume = {42},
       number = {2},
          eid = {25},
        pages = {25},
          doi = {10.1007/s12036-020-09680-5},
archivePrefix = {arXiv},
       eprint = {2101.06377},
 primaryClass = {astro-ph.IM},
       adsurl = {https://ui.adsabs.harvard.edu/abs/2021JApA...42...25J}
}

@inproceedings{singh2014astrosat,
       author = {{Singh}, Kulinder Pal and {Tandon}, S.~N. and {Agrawal}, P.~C. and {Antia}, H.~M. and {Manchanda}, R.~K. and {Yadav}, J.~S. and {Seetha}, S. and {Ramadevi}, M.~C. and {Rao}, A.~R. and {Bhattacharya}, D. and {Paul}, B. and {Sreekumar}, P. and {Bhattacharyya}, S. and {Stewart}, G.~C. and {Hutchings}, J. and {Annapurni}, S.~A. and {Ghosh}, S.~K. and {Murthy}, J. and {Pati}, A. and {Rao}, N.~K. and {Stalin}, C.~S. and {Girish}, V. and {Sankarasubramanian}, K. and {Vadawale}, S. and {Bhalerao}, V.~B. and {Dewangan}, G.~C. and {Dedhia}, D.~K. and {Hingar}, M.~K. and {Katoch}, T.~B. and {Kothare}, A.~T. and {Mirza}, I. and {Mukerjee}, K. and {Shah}, H. and {Shah}, P. and {Mohan}, R. and {Sangal}, A.~K. and {Nagabhusana}, S. and {Sriram}, S. and {Malkar}, J.~P. and {Sreekumar}, S. and {Abbey}, A.~F. and {Hansford}, G.~M. and {Beardmore}, A.~P. and {Sharma}, M.~R. and {Murthy}, S. and {Kulkarni}, R. and {Meena}, G. and {Babu}, V.~C. and {Postma}, J.},
        title = "{ASTROSAT mission}",
    booktitle = {Space Telescopes and Instrumentation 2014: Ultraviolet to Gamma Ray},
         year = 2014,
       editor = {{Takahashi}, Tadayuki and {den Herder}, Jan-Willem A. and {Bautz}, Mark},
       series = {Society of Photo-Optical Instrumentation Engineers (SPIE) Conference Series},
       volume = {9144},
        month = jul,
          eid = {91441S},
        pages = {91441S},
          doi = {10.1117/12.2062667},
       adsurl = {https://ui.adsabs.harvard.edu/abs/2014SPIE.9144E..1SS}
}

@article{agrawal2017astrosat,
  title={AstroSat: From Inception to Realization and Launch},
  author={Agrawal, PC},
  journal={Journal of Astrophysics and Astronomy},
  volume={38},
  pages={1--8},
  year={2017},
  publisher={Springer}
}

@inproceedings{nahor1993degradation,
  title={Degradation of TAUVEX optical system performance due to contamination by outgassed spacecraft materials},
  author={Nahor, Gad and Baer, Michael G and Anholt, Micha and Murat, Michael and Noter, Yoram and Lifshitz, Yeshayahu and Saar, Nachman and Braun, Ofer},
  booktitle={8th Meeting on Optical Engineering in Israel: Optical Engineering and Remote Sensing},
  volume={1971},
  pages={288--303},
  year={1993},
  organization={SPIE}
}

@inproceedings{banks2004low,
  title={Low earth orbital atomic oxygen interactions with materials},
  author={Banks, Bruce and Miller, Sharon and de Groh, Kim},
  booktitle={2nd International Energy Conversion Engineering Conference},
  pages={5638},
  year={2004}
}

@article{garoli2020mirrors,
  title={Mirrors for space telescopes: Degradation issues},
  author={Garoli, Denis and Rodriguez De Marcos, Luis V and Larruquert, Juan I and Corso, Alain J and Proietti Zaccaria, Remo and Pelizzo, Maria G},
  journal={Applied Sciences},
  volume={10},
  number={21},
  pages={7538},
  year={2020},
  publisher={MDPI}
}

@article{wenger2000simbad,
  title={The SIMBAD astronomical database-The CDS reference database for astronomical objects},
  author={Wenger, Marc and Ochsenbein, Fran{\c{c}}ois and Egret, Daniel and Dubois, Pascal and Bonnarel, Fran{\c{c}}ois and Borde, Suzanne and Genova, Fran{\c{c}}oise and Jasniewicz, G{\'e}rard and Lalo{\"e}, Suzanne and Lesteven, Soizick and others},
  journal={Astronomy and Astrophysics Supplement Series},
  volume={143},
  number={1},
  pages={9--22},
  year={2000},
  publisher={EDP Sciences}
}

@article{lang2010astrometry,
  title={Astrometry. net: Blind astrometric calibration of arbitrary astronomical images},
  author={Lang, Dustin and Hogg, David W and Mierle, Keir and Blanton, Michael and Roweis, Sam},
  journal={The astronomical journal},
  volume={139},
  number={5},
  pages={1782},
  year={2010},
  publisher={IOP Publishing}
}

@article{collaboration2016description,
       author = {{Gaia Collaboration} and {Prusti}, T. and {de Bruijne}, J.~H.~J. and {Brown}, A.~G.~A. and {Vallenari}, A. and {Babusiaux}, C. and {Bailer-Jones}, C.~A.~L. and {Bastian}, U. and {Biermann}, M. and {Evans}, D.~W. and {Eyer}, L. and {Jansen}, F. and {Jordi}, C. and {Klioner}, S.~A. and {Lammers}, U. and {Lindegren}, L. and {Luri}, X. and {Mignard}, F. and {Milligan}, D.~J. and {Panem}, C. and {Poinsignon}, V. and {Pourbaix}, D. and {Randich}, S. and {Sarri}, G. and {Sartoretti}, P. and {Siddiqui}, H.~I. and {Soubiran}, C. and {Valette}, V. and {van Leeuwen}, F. and {Walton}, N.~A. and {Aerts}, C. and {Arenou}, F. and {Cropper}, M. and {Drimmel}, R. and {H{\o}g}, E. and {Katz}, D. and {Lattanzi}, M.~G. and {O'Mullane}, W. and {Grebel}, E.~K. and {Holland}, A.~D. and {Huc}, C. and {Passot}, X. and {Bramante}, L. and {Cacciari}, C. and {Casta{\~n}eda}, J. and {Chaoul}, L. and {Cheek}, N. and {De Angeli}, F. and {Fabricius}, C. and {Guerra}, R. and {Hern{\'a}ndez}, J. and {Jean-Antoine-Piccolo}, A. and {Masana}, E. and {Messineo}, R. and {Mowlavi}, N. and {Nienartowicz}, K. and {Ord{\'o}{\~n}ez-Blanco}, D. and {Panuzzo}, P. and {Portell}, J. and {Richards}, P.~J. and {Riello}, M. and {Seabroke}, G.~M. and {Tanga}, P. and {Th{\'e}venin}, F. and {Torra}, J. and {Els}, S.~G. and {Gracia-Abril}, G. and {Comoretto}, G. and {Garcia-Reinaldos}, M. and {Lock}, T. and {Mercier}, E. and {Altmann}, M. and {Andrae}, R. and {Astraatmadja}, T.~L. and {Bellas-Velidis}, I. and {Benson}, K. and {Berthier}, J. and {Blomme}, R. and {Busso}, G. and {Carry}, B. and {Cellino}, A. and {Clementini}, G. and {Cowell}, S. and {Creevey}, O. and {Cuypers}, J. and {Davidson}, M. and {De Ridder}, J. and {de Torres}, A. and {Delchambre}, L. and {Dell'Oro}, A. and {Ducourant}, C. and {Fr{\'e}mat}, Y. and {Garc{\'\i}a-Torres}, M. and {Gosset}, E. and {Halbwachs}, J. -L. and {Hambly}, N.~C. and {Harrison}, D.~L. and {Hauser}, M. and {Hestroffer}, D. and {Hodgkin}, S.~T. and {Huckle}, H.~E. and {Hutton}, A. and {Jasniewicz}, G. and {Jordan}, S. and {Kontizas}, M. and {Korn}, A.~J. and {Lanzafame}, A.~C. and {Manteiga}, M. and {Moitinho}, A. and {Muinonen}, K. and {Osinde}, J. and {Pancino}, E. and {Pauwels}, T. and {Petit}, J. -M. and {Recio-Blanco}, A. and {Robin}, A.~C. and {Sarro}, L.~M. and {Siopis}, C. and {Smith}, M. and {Smith}, K.~W. and {Sozzetti}, A. and {Thuillot}, W. and {van Reeven}, W. and {Viala}, Y. and {Abbas}, U. and {Abreu Aramburu}, A. and {Accart}, S. and {Aguado}, J.~J. and {Allan}, P.~M. and {Allasia}, W. and {Altavilla}, G. and {{\'A}lvarez}, M.~A. and {Alves}, J. and {Anderson}, R.~I. and {Andrei}, A.~H. and {Anglada Varela}, E. and {Antiche}, E. and {Antoja}, T. and {Ant{\'o}n}, S. and {Arcay}, B. and {Atzei}, A. and {Ayache}, L. and {Bach}, N. and {Baker}, S.~G. and {Balaguer-N{\'u}{\~n}ez}, L. and {Barache}, C. and {Barata}, C. and {Barbier}, A. and {Barblan}, F. and {Baroni}, M. and {Barrado y Navascu{\'e}s}, D. and {Barros}, M. and {Barstow}, M.~A. and {Becciani}, U. and {Bellazzini}, M. and {Bellei}, G. and {Bello Garc{\'\i}a}, A. and {Belokurov}, V. and {Bendjoya}, P. and {Berihuete}, A. and {Bianchi}, L. and {Bienaym{\'e}}, O. and {Billebaud}, F. and {Blagorodnova}, N. and {Blanco-Cuaresma}, S. and {Boch}, T. and {Bombrun}, A. and {Borrachero}, R. and {Bouquillon}, S. and {Bourda}, G. and {Bouy}, H. and {Bragaglia}, A. and {Breddels}, M.~A. and {Brouillet}, N. and {Br{\"u}semeister}, T. and {Bucciarelli}, B. and {Budnik}, F. and {Burgess}, P. and {Burgon}, R. and {Burlacu}, A. and {Busonero}, D. and {Buzzi}, R. and {Caffau}, E. and {Cambras}, J. and {Campbell}, H. and {Cancelliere}, R. and {Cantat-Gaudin}, T. and {Carlucci}, T. and {Carrasco}, J.~M. and {Castellani}, M. and {Charlot}, P. and {Charnas}, J. and {Charvet}, P. and {Chassat}, F. and {Chiavassa}, A. and {Clotet}, M. and {Cocozza}, G. and {Collins}, R.~S. and {Collins}, P. and {Costigan}, G. and {Crifo}, F. and {Cross}, N.~J.~G. and {Crosta}, M. and {Crowley}, C. and {Dafonte}, C. and {Damerdji}, Y. and {Dapergolas}, A. and {David}, P. and {David}, M. and {De Cat}, P. and {de Felice}, F. and {de Laverny}, P. and {De Luise}, F. and {De March}, R. and {de Martino}, D. and {de Souza}, R. and {Debosscher}, J. and {del Pozo}, E. and {Delbo}, M. and {Delgado}, A. and {Delgado}, H.~E. and {di Marco}, F. and {Di Matteo}, P. and {Diakite}, S. and {Distefano}, E. and {Dolding}, C. and {Dos Anjos}, S. and {Drazinos}, P. and {Dur{\'a}n}, J. and {Dzigan}, Y. and {Ecale}, E. and {Edvardsson}, B. and {Enke}, H. and {Erdmann}, M. and {Escolar}, D. and {Espina}, M. and {Evans}, N.~W. and {Eynard Bontemps}, G. and {Fabre}, C. and {Fabrizio}, M. and {Faigler}, S. and {Falc{\~a}o}, A.~J. and {Farr{\`a}s Casas}, M. and {Faye}, F. and {Federici}, L. and {Fedorets}, G. and {Fern{\'a}ndez-Hern{\'a}ndez}, J. and {Fernique}, P. and {Fienga}, A. and {Figueras}, F. and {Filippi}, F. and {Findeisen}, K. and {Fonti}, A. and {Fouesneau}, M. and {Fraile}, E. and {Fraser}, M. and {Fuchs}, J. and {Furnell}, R. and {Gai}, M. and {Galleti}, S. and {Galluccio}, L. and {Garabato}, D. and {Garc{\'\i}a-Sedano}, F. and {Gar{\'e}}, P. and {Garofalo}, A. and {Garralda}, N. and {Gavras}, P. and {Gerssen}, J. and {Geyer}, R. and {Gilmore}, G. and {Girona}, S. and {Giuffrida}, G. and {Gomes}, M. and {Gonz{\'a}lez-Marcos}, A. and {Gonz{\'a}lez-N{\'u}{\~n}ez}, J. and {Gonz{\'a}lez-Vidal}, J.~J. and {Granvik}, M. and {Guerrier}, A. and {Guillout}, P. and {Guiraud}, J. and {G{\'u}rpide}, A. and {Guti{\'e}rrez-S{\'a}nchez}, R. and {Guy}, L.~P. and {Haigron}, R. and {Hatzidimitriou}, D. and {Haywood}, M. and {Heiter}, U. and {Helmi}, A. and {Hobbs}, D. and {Hofmann}, W. and {Holl}, B. and {Holland}, G. and {Hunt}, J.~A.~S. and {Hypki}, A. and {Icardi}, V. and {Irwin}, M. and {Jevardat de Fombelle}, G. and {Jofr{\'e}}, P. and {Jonker}, P.~G. and {Jorissen}, A. and {Julbe}, F. and {Karampelas}, A. and {Kochoska}, A. and {Kohley}, R. and {Kolenberg}, K. and {Kontizas}, E. and {Koposov}, S.~E. and {Kordopatis}, G. and {Koubsky}, P. and {Kowalczyk}, A. and {Krone-Martins}, A. and {Kudryashova}, M. and {Kull}, I. and {Bachchan}, R.~K. and {Lacoste-Seris}, F. and {Lanza}, A.~F. and {Lavigne}, J. -B. and {Le Poncin-Lafitte}, C. and {Lebreton}, Y. and {Lebzelter}, T. and {Leccia}, S. and {Leclerc}, N. and {Lecoeur-Taibi}, I. and {Lemaitre}, V. and {Lenhardt}, H. and {Leroux}, F. and {Liao}, S. and {Licata}, E. and {Lindstr{\o}m}, H.~E.~P. and {Lister}, T.~A. and {Livanou}, E. and {Lobel}, A. and {L{\"o}ffler}, W. and {L{\'o}pez}, M. and {Lopez-Lozano}, A. and {Lorenz}, D. and {Loureiro}, T. and {MacDonald}, I. and {Magalh{\~a}es Fernandes}, T. and {Managau}, S. and {Mann}, R.~G. and {Mantelet}, G. and {Marchal}, O. and {Marchant}, J.~M. and {Marconi}, M. and {Marie}, J. and {Marinoni}, S. and {Marrese}, P.~M. and {Marschalk{\'o}}, G. and {Marshall}, D.~J. and {Mart{\'\i}n-Fleitas}, J.~M. and {Martino}, M. and {Mary}, N. and {Matijevi{\v{c}}}, G. and {Mazeh}, T. and {McMillan}, P.~J. and {Messina}, S. and {Mestre}, A. and {Michalik}, D. and {Millar}, N.~R. and {Miranda}, B.~M.~H. and {Molina}, D. and {Molinaro}, R. and {Molinaro}, M. and {Moln{\'a}r}, L. and {Moniez}, M. and {Montegriffo}, P. and {Monteiro}, D. and {Mor}, R. and {Mora}, A. and {Morbidelli}, R. and {Morel}, T. and {Morgenthaler}, S. and {Morley}, T. and {Morris}, D. and {Mulone}, A.~F. and {Muraveva}, T. and {Musella}, I. and {Narbonne}, J. and {Nelemans}, G. and {Nicastro}, L. and {Noval}, L. and {Ord{\'e}novic}, C. and {Ordieres-Mer{\'e}}, J. and {Osborne}, P. and {Pagani}, C. and {Pagano}, I. and {Pailler}, F. and {Palacin}, H. and {Palaversa}, L. and {Parsons}, P. and {Paulsen}, T. and {Pecoraro}, M. and {Pedrosa}, R. and {Pentik{\"a}inen}, H. and {Pereira}, J. and {Pichon}, B. and {Piersimoni}, A.~M. and {Pineau}, F. -X. and {Plachy}, E. and {Plum}, G. and {Poujoulet}, E. and {Pr{\v{s}}a}, A. and {Pulone}, L. and {Ragaini}, S. and {Rago}, S. and {Rambaux}, N. and {Ramos-Lerate}, M. and {Ranalli}, P. and {Rauw}, G. and {Read}, A. and {Regibo}, S. and {Renk}, F. and {Reyl{\'e}}, C. and {Ribeiro}, R.~A. and {Rimoldini}, L. and {Ripepi}, V. and {Riva}, A. and {Rixon}, G. and {Roelens}, M. and {Romero-G{\'o}mez}, M. and {Rowell}, N. and {Royer}, F. and {Rudolph}, A. and {Ruiz-Dern}, L. and {Sadowski}, G. and {Sagrist{\`a} Sell{\'e}s}, T. and {Sahlmann}, J. and {Salgado}, J. and {Salguero}, E. and {Sarasso}, M. and {Savietto}, H. and {Schnorhk}, A. and {Schultheis}, M. and {Sciacca}, E. and {Segol}, M. and {Segovia}, J.~C. and {Segransan}, D. and {Serpell}, E. and {Shih}, I. -C. and {Smareglia}, R. and {Smart}, R.~L. and {Smith}, C. and {Solano}, E. and {Solitro}, F. and {Sordo}, R. and {Soria Nieto}, S. and {Souchay}, J. and {Spagna}, A. and {Spoto}, F. and {Stampa}, U. and {Steele}, I.~A. and {Steidelm{\"u}ller}, H. and {Stephenson}, C.~A. and {Stoev}, H. and {Suess}, F.~F. and {S{\"u}veges}, M. and {Surdej}, J. and {Szabados}, L. and {Szegedi-Elek}, E. and {Tapiador}, D. and {Taris}, F. and {Tauran}, G. and {Taylor}, M.~B. and {Teixeira}, R. and {Terrett}, D. and {Tingley}, B. and {Trager}, S.~C. and {Turon}, C. and {Ulla}, A. and {Utrilla}, E. and {Valentini}, G. and {van Elteren}, A. and {Van Hemelryck}, E. and {van Leeuwen}, M. and {Varadi}, M. and {Vecchiato}, A. and {Veljanoski}, J. and {Via}, T. and {Vicente}, D. and {Vogt}, S. and {Voss}, H. and {Votruba}, V. and {Voutsinas}, S. and {Walmsley}, G. and {Weiler}, M. and {Weingrill}, K. and {Werner}, D. and {Wevers}, T. and {Whitehead}, G. and {Wyrzykowski}, {\L}. and {Yoldas}, A. and {{\v{Z}}erjal}, M. and {Zucker}, S. and {Zurbach}, C. and {Zwitter}, T. and {Alecu}, A. and {Allen}, M. and {Allende Prieto}, C. and {Amorim}, A. and {Anglada-Escud{\'e}}, G. and {Arsenijevic}, V. and {Azaz}, S. and {Balm}, P. and {Beck}, M. and {Bernstein}, H. -H. and {Bigot}, L. and {Bijaoui}, A. and {Blasco}, C. and {Bonfigli}, M. and {Bono}, G. and {Boudreault}, S. and {Bressan}, A. and {Brown}, S. and {Brunet}, P. -M. and {Bunclark}, P. and {Buonanno}, R. and {Butkevich}, A.~G. and {Carret}, C. and {Carrion}, C. and {Chemin}, L. and {Ch{\'e}reau}, F. and {Corcione}, L. and {Darmigny}, E. and {de Boer}, K.~S. and {de Teodoro}, P. and {de Zeeuw}, P.~T. and {Delle Luche}, C. and {Domingues}, C.~D. and {Dubath}, P. and {Fodor}, F. and {Fr{\'e}zouls}, B. and {Fries}, A. and {Fustes}, D. and {Fyfe}, D. and {Gallardo}, E. and {Gallegos}, J. and {Gardiol}, D. and {Gebran}, M. and {Gomboc}, A. and {G{\'o}mez}, A. and {Grux}, E. and {Gueguen}, A. and {Heyrovsky}, A. and {Hoar}, J. and {Iannicola}, G. and {Isasi Parache}, Y. and {Janotto}, A. -M. and {Joliet}, E. and {Jonckheere}, A. and {Keil}, R. and {Kim}, D. -W. and {Klagyivik}, P. and {Klar}, J. and {Knude}, J. and {Kochukhov}, O. and {Kolka}, I. and {Kos}, J. and {Kutka}, A. and {Lainey}, V. and {LeBouquin}, D. and {Liu}, C. and {Loreggia}, D. and {Makarov}, V.~V. and {Marseille}, M.~G. and {Martayan}, C. and {Martinez-Rubi}, O. and {Massart}, B. and {Meynadier}, F. and {Mignot}, S. and {Munari}, U. and {Nguyen}, A. -T. and {Nordlander}, T. and {Ocvirk}, P. and {O'Flaherty}, K.~S. and {Olias Sanz}, A. and {Ortiz}, P. and {Osorio}, J. and {Oszkiewicz}, D. and {Ouzounis}, A. and {Palmer}, M. and {Park}, P. and {Pasquato}, E. and {Peltzer}, C. and {Peralta}, J. and {P{\'e}turaud}, F. and {Pieniluoma}, T. and {Pigozzi}, E. and {Poels}, J. and {Prat}, G. and {Prod'homme}, T. and {Raison}, F. and {Rebordao}, J.~M. and {Risquez}, D. and {Rocca-Volmerange}, B. and {Rosen}, S. and {Ruiz-Fuertes}, M.~I. and {Russo}, F. and {Sembay}, S. and {Serraller Vizcaino}, I. and {Short}, A. and {Siebert}, A. and {Silva}, H. and {Sinachopoulos}, D. and {Slezak}, E. and {Soffel}, M. and {Sosnowska}, D. and {Strai{\v{z}}ys}, V. and {ter Linden}, M. and {Terrell}, D. and {Theil}, S. and {Tiede}, C. and {Troisi}, L. and {Tsalmantza}, P. and {Tur}, D. and {Vaccari}, M. and {Vachier}, F. and {Valles}, P. and {Van Hamme}, W. and {Veltz}, L. and {Virtanen}, J. and {Wallut}, J. -M. and {Wichmann}, R. and {Wilkinson}, M.~I. and {Ziaeepour}, H. and {Zschocke}, S.},
        title = "{The Gaia mission}",
      journal = {\aap},
         year = 2016,
        month = nov,
       volume = {595},
          eid = {A1},
        pages = {A1},
          doi = {10.1051/0004-6361/201629272},
archivePrefix = {arXiv},
       eprint = {1609.04153},
 primaryClass = {astro-ph.IM},
       adsurl = {https://ui.adsabs.harvard.edu/abs/2016A&A...595A...1G}
}

@ARTICLE{2023A&A...674A...1G,
       author = {{Gaia Collaboration} and {Vallenari}, A. and {Brown}, A.~G.~A. and {Prusti}, T. and {de Bruijne}, J.~H.~J. and {Arenou}, F. and {Babusiaux}, C. and {Biermann}, M. and {Creevey}, O.~L. and {Ducourant}, C. and {Evans}, D.~W. and {Eyer}, L. and {Guerra}, R. and {Hutton}, A. and {Jordi}, C. and {Klioner}, S.~A. and {Lammers}, U.~L. and {Lindegren}, L. and {Luri}, X. and {Mignard}, F. and {Panem}, C. and {Pourbaix}, D. and {Randich}, S. and {Sartoretti}, P. and {Soubiran}, C. and {Tanga}, P. and {Walton}, N.~A. and {Bailer-Jones}, C.~A.~L. and {Bastian}, U. and {Drimmel}, R. and {Jansen}, F. and {Katz}, D. and {Lattanzi}, M.~G. and {van Leeuwen}, F. and {Bakker}, J. and {Cacciari}, C. and {Casta{\~n}eda}, J. and {De Angeli}, F. and {Fabricius}, C. and {Fouesneau}, M. and {Fr{\'e}mat}, Y. and {Galluccio}, L. and {Guerrier}, A. and {Heiter}, U. and {Masana}, E. and {Messineo}, R. and {Mowlavi}, N. and {Nicolas}, C. and {Nienartowicz}, K. and {Pailler}, F. and {Panuzzo}, P. and {Riclet}, F. and {Roux}, W. and {Seabroke}, G.~M. and {Sordo}, R. and {Th{\'e}venin}, F. and {Gracia-Abril}, G. and {Portell}, J. and {Teyssier}, D. and {Altmann}, M. and {Andrae}, R. and {Audard}, M. and {Bellas-Velidis}, I. and {Benson}, K. and {Berthier}, J. and {Blomme}, R. and {Burgess}, P.~W. and {Busonero}, D. and {Busso}, G. and {C{\'a}novas}, H. and {Carry}, B. and {Cellino}, A. and {Cheek}, N. and {Clementini}, G. and {Damerdji}, Y. and {Davidson}, M. and {de Teodoro}, P. and {Nu{\~n}ez Campos}, M. and {Delchambre}, L. and {Dell'Oro}, A. and {Esquej}, P. and {Fern{\'a}ndez-Hern{\'a}ndez}, J. and {Fraile}, E. and {Garabato}, D. and {Garc{\'\i}a-Lario}, P. and {Gosset}, E. and {Haigron}, R. and {Halbwachs}, J. -L. and {Hambly}, N.~C. and {Harrison}, D.~L. and {Hern{\'a}ndez}, J. and {Hestroffer}, D. and {Hodgkin}, S.~T. and {Holl}, B. and {Jan{\ss}en}, K. and {Jevardat de Fombelle}, G. and {Jordan}, S. and {Krone-Martins}, A. and {Lanzafame}, A.~C. and {L{\"o}ffler}, W. and {Marchal}, O. and {Marrese}, P.~M. and {Moitinho}, A. and {Muinonen}, K. and {Osborne}, P. and {Pancino}, E. and {Pauwels}, T. and {Recio-Blanco}, A. and {Reyl{\'e}}, C. and {Riello}, M. and {Rimoldini}, L. and {Roegiers}, T. and {Rybizki}, J. and {Sarro}, L.~M. and {Siopis}, C. and {Smith}, M. and {Sozzetti}, A. and {Utrilla}, E. and {van Leeuwen}, M. and {Abbas}, U. and {{\'A}brah{\'a}m}, P. and {Abreu Aramburu}, A. and {Aerts}, C. and {Aguado}, J.~J. and {Ajaj}, M. and {Aldea-Montero}, F. and {Altavilla}, G. and {{\'A}lvarez}, M.~A. and {Alves}, J. and {Anders}, F. and {Anderson}, R.~I. and {Anglada Varela}, E. and {Antoja}, T. and {Baines}, D. and {Baker}, S.~G. and {Balaguer-N{\'u}{\~n}ez}, L. and {Balbinot}, E. and {Balog}, Z. and {Barache}, C. and {Barbato}, D. and {Barros}, M. and {Barstow}, M.~A. and {Bartolom{\'e}}, S. and {Bassilana}, J. -L. and {Bauchet}, N. and {Becciani}, U. and {Bellazzini}, M. and {Berihuete}, A. and {Bernet}, M. and {Bertone}, S. and {Bianchi}, L. and {Binnenfeld}, A. and {Blanco-Cuaresma}, S. and {Blazere}, A. and {Boch}, T. and {Bombrun}, A. and {Bossini}, D. and {Bouquillon}, S. and {Bragaglia}, A. and {Bramante}, L. and {Breedt}, E. and {Bressan}, A. and {Brouillet}, N. and {Brugaletta}, E. and {Bucciarelli}, B. and {Burlacu}, A. and {Butkevich}, A.~G. and {Buzzi}, R. and {Caffau}, E. and {Cancelliere}, R. and {Cantat-Gaudin}, T. and {Carballo}, R. and {Carlucci}, T. and {Carnerero}, M.~I. and {Carrasco}, J.~M. and {Casamiquela}, L. and {Castellani}, M. and {Castro-Ginard}, A. and {Chaoul}, L. and {Charlot}, P. and {Chemin}, L. and {Chiaramida}, V. and {Chiavassa}, A. and {Chornay}, N. and {Comoretto}, G. and {Contursi}, G. and {Cooper}, W.~J. and {Cornez}, T. and {Cowell}, S. and {Crifo}, F. and {Cropper}, M. and {Crosta}, M. and {Crowley}, C. and {Dafonte}, C. and {Dapergolas}, A. and {David}, M. and {David}, P. and {de Laverny}, P. and {De Luise}, F. and {De March}, R. and {De Ridder}, J. and {de Souza}, R. and {de Torres}, A. and {del Peloso}, E.~F. and {del Pozo}, E. and {Delbo}, M. and {Delgado}, A. and {Delisle}, J. -B. and {Demouchy}, C. and {Dharmawardena}, T.~E. and {Di Matteo}, P. and {Diakite}, S. and {Diener}, C. and {Distefano}, E. and {Dolding}, C. and {Edvardsson}, B. and {Enke}, H. and {Fabre}, C. and {Fabrizio}, M. and {Faigler}, S. and {Fedorets}, G. and {Fernique}, P. and {Fienga}, A. and {Figueras}, F. and {Fournier}, Y. and {Fouron}, C. and {Fragkoudi}, F. and {Gai}, M. and {Garcia-Gutierrez}, A. and {Garcia-Reinaldos}, M. and {Garc{\'\i}a-Torres}, M. and {Garofalo}, A. and {Gavel}, A. and {Gavras}, P. and {Gerlach}, E. and {Geyer}, R. and {Giacobbe}, P. and {Gilmore}, G. and {Girona}, S. and {Giuffrida}, G. and {Gomel}, R. and {Gomez}, A. and {Gonz{\'a}lez-N{\'u}{\~n}ez}, J. and {Gonz{\'a}lez-Santamar{\'\i}a}, I. and {Gonz{\'a}lez-Vidal}, J.~J. and {Granvik}, M. and {Guillout}, P. and {Guiraud}, J. and {Guti{\'e}rrez-S{\'a}nchez}, R. and {Guy}, L.~P. and {Hatzidimitriou}, D. and {Hauser}, M. and {Haywood}, M. and {Helmer}, A. and {Helmi}, A. and {Sarmiento}, M.~H. and {Hidalgo}, S.~L. and {Hilger}, T. and {H{\l}adczuk}, N. and {Hobbs}, D. and {Holland}, G. and {Huckle}, H.~E. and {Jardine}, K. and {Jasniewicz}, G. and {Jean-Antoine Piccolo}, A. and {Jim{\'e}nez-Arranz}, {\'O}. and {Jorissen}, A. and {Juaristi Campillo}, J. and {Julbe}, F. and {Karbevska}, L. and {Kervella}, P. and {Khanna}, S. and {Kontizas}, M. and {Kordopatis}, G. and {Korn}, A.~J. and {K{\'o}sp{\'a}l}, {\'A}. and {Kostrzewa-Rutkowska}, Z. and {Kruszy{\'n}ska}, K. and {Kun}, M. and {Laizeau}, P. and {Lambert}, S. and {Lanza}, A.~F. and {Lasne}, Y. and {Le Campion}, J. -F. and {Lebreton}, Y. and {Lebzelter}, T. and {Leccia}, S. and {Leclerc}, N. and {Lecoeur-Taibi}, I. and {Liao}, S. and {Licata}, E.~L. and {Lindstr{\o}m}, H.~E.~P. and {Lister}, T.~A. and {Livanou}, E. and {Lobel}, A. and {Lorca}, A. and {Loup}, C. and {Madrero Pardo}, P. and {Magdaleno Romeo}, A. and {Managau}, S. and {Mann}, R.~G. and {Manteiga}, M. and {Marchant}, J.~M. and {Marconi}, M. and {Marcos}, J. and {Marcos Santos}, M.~M.~S. and {Mar{\'\i}n Pina}, D. and {Marinoni}, S. and {Marocco}, F. and {Marshall}, D.~J. and {Martin Polo}, L. and {Mart{\'\i}n-Fleitas}, J.~M. and {Marton}, G. and {Mary}, N. and {Masip}, A. and {Massari}, D. and {Mastrobuono-Battisti}, A. and {Mazeh}, T. and {McMillan}, P.~J. and {Messina}, S. and {Michalik}, D. and {Millar}, N.~R. and {Mints}, A. and {Molina}, D. and {Molinaro}, R. and {Moln{\'a}r}, L. and {Monari}, G. and {Mongui{\'o}}, M. and {Montegriffo}, P. and {Montero}, A. and {Mor}, R. and {Mora}, A. and {Morbidelli}, R. and {Morel}, T. and {Morris}, D. and {Muraveva}, T. and {Murphy}, C.~P. and {Musella}, I. and {Nagy}, Z. and {Noval}, L. and {Oca{\~n}a}, F. and {Ogden}, A. and {Ordenovic}, C. and {Osinde}, J.~O. and {Pagani}, C. and {Pagano}, I. and {Palaversa}, L. and {Palicio}, P.~A. and {Pallas-Quintela}, L. and {Panahi}, A. and {Payne-Wardenaar}, S. and {Pe{\~n}alosa Esteller}, X. and {Penttil{\"a}}, A. and {Pichon}, B. and {Piersimoni}, A.~M. and {Pineau}, F. -X. and {Plachy}, E. and {Plum}, G. and {Poggio}, E. and {Pr{\v{s}}a}, A. and {Pulone}, L. and {Racero}, E. and {Ragaini}, S. and {Rainer}, M. and {Raiteri}, C.~M. and {Rambaux}, N. and {Ramos}, P. and {Ramos-Lerate}, M. and {Re Fiorentin}, P. and {Regibo}, S. and {Richards}, P.~J. and {Rios Diaz}, C. and {Ripepi}, V. and {Riva}, A. and {Rix}, H. -W. and {Rixon}, G. and {Robichon}, N. and {Robin}, A.~C. and {Robin}, C. and {Roelens}, M. and {Rogues}, H.~R.~O. and {Rohrbasser}, L. and {Romero-G{\'o}mez}, M. and {Rowell}, N. and {Royer}, F. and {Ruz Mieres}, D. and {Rybicki}, K.~A. and {Sadowski}, G. and {S{\'a}ez N{\'u}{\~n}ez}, A. and {Sagrist{\`a} Sell{\'e}s}, A. and {Sahlmann}, J. and {Salguero}, E. and {Samaras}, N. and {Sanchez Gimenez}, V. and {Sanna}, N. and {Santove{\~n}a}, R. and {Sarasso}, M. and {Schultheis}, M. and {Sciacca}, E. and {Segol}, M. and {Segovia}, J.~C. and {S{\'e}gransan}, D. and {Semeux}, D. and {Shahaf}, S. and {Siddiqui}, H.~I. and {Siebert}, A. and {Siltala}, L. and {Silvelo}, A. and {Slezak}, E. and {Slezak}, I. and {Smart}, R.~L. and {Snaith}, O.~N. and {Solano}, E. and {Solitro}, F. and {Souami}, D. and {Souchay}, J. and {Spagna}, A. and {Spina}, L. and {Spoto}, F. and {Steele}, I.~A. and {Steidelm{\"u}ller}, H. and {Stephenson}, C.~A. and {S{\"u}veges}, M. and {Surdej}, J. and {Szabados}, L. and {Szegedi-Elek}, E. and {Taris}, F. and {Taylor}, M.~B. and {Teixeira}, R. and {Tolomei}, L. and {Tonello}, N. and {Torra}, F. and {Torra}, J. and {Torralba Elipe}, G. and {Trabucchi}, M. and {Tsounis}, A.~T. and {Turon}, C. and {Ulla}, A. and {Unger}, N. and {Vaillant}, M.~V. and {van Dillen}, E. and {van Reeven}, W. and {Vanel}, O. and {Vecchiato}, A. and {Viala}, Y. and {Vicente}, D. and {Voutsinas}, S. and {Weiler}, M. and {Wevers}, T. and {Wyrzykowski}, {\L}. and {Yoldas}, A. and {Yvard}, P. and {Zhao}, H. and {Zorec}, J. and {Zucker}, S. and {Zwitter}, T.},
        title = "{Gaia Data Release 3. Summary of the content and survey properties}",
      journal = {\aap},
         year = 2023,
        month = jun,
       volume = {674},
          eid = {A1},
        pages = {A1},
          doi = {10.1051/0004-6361/202243940},
archivePrefix = {arXiv},
       eprint = {2208.00211},
 primaryClass = {astro-ph.GA},
       adsurl = {https://ui.adsabs.harvard.edu/abs/2023A&A...674A...1G}
}

@Article{Hunter:2007,
  Author    = {Hunter, J. D.},
  Title     = {Matplotlib: A 2D graphics environment},
  Journal   = {Computing in Science \& Engineering},
  Volume    = {9},
  Number    = {3},
  Pages     = {90--95},
  publisher = {IEEE COMPUTER SOC},
  doi       = {10.1109/MCSE.2007.55},
  year      = 2007
}

@Article{         harris2020array,
 title         = {Array programming with {NumPy}},
 author        = {Charles R. Harris and K. Jarrod Millman and St{\'{e}}fan J.
                 van der Walt and Ralf Gommers and Pauli Virtanen and David
                 Cournapeau and Eric Wieser and Julian Taylor and Sebastian
                 Berg and Nathaniel J. Smith and Robert Kern and Matti Picus
                 and Stephan Hoyer and Marten H. van Kerkwijk and Matthew
                 Brett and Allan Haldane and Jaime Fern{\'{a}}ndez del
                 R{\'{i}}o and Mark Wiebe and Pearu Peterson and Pierre
                 G{\'{e}}rard-Marchant and Kevin Sheppard and Tyler Reddy and
                 Warren Weckesser and Hameer Abbasi and Christoph Gohlke and
                 Travis E. Oliphant},
 year          = {2020},
 month         = sep,
 journal       = {Nature},
 volume        = {585},
 number        = {7825},
 pages         = {357--362},
 doi           = {10.1038/s41586-020-2649-2},
 publisher     = {Springer Science and Business Media {LLC}},
 url           = {https://doi.org/10.1038/s41586-020-2649-2}
}

@article{prajwel_aafitrans,
    author = {Prajwel Joseph},
    title = {aafitrans: v0.2.0},
    publisher = {Zenodo},
    year = 2023,
    journal = "",
    doi = {10.5281/zenodo.10041152},
    url = {https://doi.org/10.5281/zenodo.10041152}
}

@ARTICLE{2020SciPy-NMeth,
  author  = {Virtanen, Pauli and Gommers, Ralf and Oliphant, Travis E. and
            Haberland, Matt and Reddy, Tyler and Cournapeau, David and
            Burovski, Evgeni and Peterson, Pearu and Weckesser, Warren and
            Bright, Jonathan and {van der Walt}, St{\'e}fan J. and
            Brett, Matthew and Wilson, Joshua and Millman, K. Jarrod and
            Mayorov, Nikolay and Nelson, Andrew R. J. and Jones, Eric and
            Kern, Robert and Larson, Eric and Carey, C J and
            Polat, {\.I}lhan and Feng, Yu and Moore, Eric W. and
            {VanderPlas}, Jake and Laxalde, Denis and Perktold, Josef and
            Cimrman, Robert and Henriksen, Ian and Quintero, E. A. and
            Harris, Charles R. and Archibald, Anne M. and
            Ribeiro, Ant{\^o}nio H. and Pedregosa, Fabian and
            {van Mulbregt}, Paul and {SciPy 1.0 Contributors}},
  title   = {{{SciPy} 1.0: Fundamental Algorithms for Scientific
            Computing in Python}},
  journal = {Nature Methods},
  year    = {2020},
  volume  = {17},
  pages   = {261--272},
  adsurl  = {https://rdcu.be/b08Wh},
  doi     = {10.1038/s41592-019-0686-2},
}

@software{reback2020pandas,
    author       = {The pandas development team},
    title        = {pandas-dev/pandas: Pandas},
    month        = feb,
    year         = 2020,
    publisher    = {Zenodo},
    version      = {1.1.3},
    doi          = {10.5281/zenodo.3509134},
    url          = {https://doi.org/10.5281/zenodo.3509134}
}

@ARTICLE{2005ApJ...622..759G,
       author = {{G{\'o}rski}, K.~M. and {Hivon}, E. and {Banday}, A.~J. and {Wandelt}, B.~D. and {Hansen}, F.~K. and {Reinecke}, M. and {Bartelmann}, M.},
        title = "{HEALPix: A Framework for High-Resolution Discretization and Fast Analysis of Data Distributed on the Sphere}",
      journal = {\apj},
         year = 2005,
        month = apr,
       volume = {622},
       number = {2},
        pages = {759-771},
          doi = {10.1086/427976},
archivePrefix = {arXiv},
       eprint = {astro-ph/0409513},
 primaryClass = {astro-ph},
       adsurl = {https://ui.adsabs.harvard.edu/abs/2005ApJ...622..759G}
}

@INPROCEEDINGS{2003ASPC..295..489J,
       author = {{Joye}, W.~A. and {Mandel}, E.},
        title = "{New Features of SAOImage DS9}",
    booktitle = {Astronomical Data Analysis Software and Systems XII},
         year = 2003,
       editor = {{Payne}, H.~E. and {Jedrzejewski}, R.~I. and {Hook}, R.~N.},
       series = {Astronomical Society of the Pacific Conference Series},
       volume = {295},
        month = jan,
        pages = {489},
       adsurl = {https://ui.adsabs.harvard.edu/abs/2003ASPC..295..489J}
}

@INPROCEEDINGS{2005ASPC..347...29T,
       author = {{Taylor}, M.~B.},
        title = "{TOPCAT \& STIL: Starlink Table/VOTable Processing Software}",
    booktitle = {Astronomical Data Analysis Software and Systems XIV},
         year = 2005,
       editor = {{Shopbell}, P. and {Britton}, M. and {Ebert}, R.},
       series = {Astronomical Society of the Pacific Conference Series},
       volume = {347},
        month = dec,
        pages = {29},
       adsurl = {https://ui.adsabs.harvard.edu/abs/2005ASPC..347...29T}
}

@article{raybaut2009spyder,
  title={Spyder-documentation},
  author={Raybaut, Pierre},
  journal={Available online at: pythonhosted. org},
  year={2009}
}

@conference{Kluyver2016jupyter,
Title = {Jupyter Notebooks -- a publishing format for reproducible computational workflows},
Author = {Thomas Kluyver and Benjamin Ragan-Kelley and Fernando P{\'e}rez and Brian Granger and Matthias Bussonnier and Jonathan Frederic and Kyle Kelley and Jessica Hamrick and Jason Grout and Sylvain Corlay and Paul Ivanov and Dami{\'a}n Avila and Safia Abdalla and Carol Willing},
Booktitle = {Positioning and Power in Academic Publishing: Players, Agents and Agendas},
Editor = {F. Loizides and B. Schmidt},
Organization = {IOS Press},
Pages = {87 - 90},
Year = {2016}
}

@article{choi1994medians,
  title={On the medians of gamma distributions and an equation of Ramanujan},
  author={Choi, Kwok Pui},
  journal={Proceedings of the American Mathematical Society},
  volume={121},
  number={1},
  pages={245--251},
  year={1994}
}

@article{tandon2017orbit,
       author = {{Tandon}, S.~N. and {Subramaniam}, Annapurni and {Girish}, V. and {Postma}, J. and {Sankarasubramanian}, K. and {Sriram}, S. and {Stalin}, C.~S. and {Mondal}, C. and {Sahu}, S. and {Joseph}, P. and {Hutchings}, J. and {Ghosh}, S.~K. and {Barve}, I.~V. and {George}, K. and {Kamath}, P.~U. and {Kathiravan}, S. and {Kumar}, A. and {Lancelot}, J.~P. and {Leahy}, D. and {Mahesh}, P.~K. and {Mohan}, R. and {Nagabhushana}, S. and {Pati}, A.~K. and {Kameswara Rao}, N. and {Sreedhar}, Y.~H. and {Sreekumar}, P.},
        title = "{In-orbit Calibrations of the Ultraviolet Imaging Telescope}",
      journal = {\aj},
         year = 2017,
        month = sep,
       volume = {154},
       number = {3},
          eid = {128},
        pages = {128},
          doi = {10.3847/1538-3881/aa8451},
archivePrefix = {arXiv},
       eprint = {1705.03715},
 primaryClass = {astro-ph.IM},
       adsurl = {https://ui.adsabs.harvard.edu/abs/2017AJ....154..128T}
}

@dataset{dagore_2025_18076974,
  author       = {Dagore, Akanksha and
                  Joseph, Prajwel and
                  Tandon, S.N. and
                  Subramaniam, Annapurni and
                  Ghosh, S.K. and
                  Chelliah Subramonian, Stalin},
  title        = {Nine years of UVIT: assessing sensitivity
                   variation
                  },
  month        = dec,
  year         = 2025,
  publisher    = {Zenodo},
  version      = {V1.0.1},
  doi          = {10.5281/zenodo.18076974},
  url          = {https://doi.org/10.5281/zenodo.18076974},
}
\bibliographystyle{aasjournal}



\end{document}